\documentclass[12pt,a4paper]{article}
\usepackage{amsmath,amssymb,amsthm,bbm,stmaryrd}
\usepackage{graphicx,psfrag}
\graphicspath{{images/}}
\usepackage[english]{babel}
\usepackage{hyperref}

\newenvironment{proofOF}[2]{\removelastskip\vspace{6pt}\noindent
 {\it Proof #1.}~\rm#2}{\par\vspace{6pt}}

\numberwithin{equation}{section}

\newtheorem{theorem}{Theorem}[section]
\newtheorem{proposition}[theorem]{Proposition}
\newtheorem{lemma}[theorem]{Lemma}

\newcommand{\e}{\varepsilon}

\newcommand{\E}{\mathbbm{E}}
\newcommand{\Pb}{\mathbbm{P}}
\newcommand{\Or}{\mathcal{O}}
\newcommand{\R}{\mathbbm{R}}
\newcommand{\Z}{\mathbbm{Z}}
\newcommand{\N}{\mathbbm{N}}
\newcommand{\Id}{\mathbbm{1}}
\newcommand{\dx}{{\rm d}}
\newcommand{\ket}[1]{\arrowvert #1 \rangle}
\newcommand{\bra}[1]{\langle #1 \arrowvert}

\DeclareMathOperator*{\Pf}{Pf}
\DeclareMathOperator*{\Det}{Det}
\DeclareMathOperator*{\Tr}{Tr}
\DeclareMathOperator*{\Aip}{Ai^\prime}
\DeclareMathOperator*{\Ai}{Ai}
\DeclareMathOperator*{\sgn}{sgn}

\title{Polynuclear growth on a flat substrate and edge scaling of GOE eigenvalues}
\author{Patrik L. Ferrari\\Zentrum Mathematik\\Technische Universit\"at M\"unchen}
\date{\today}

\begin{document}
\maketitle

\begin{abstract}
We consider the polynuclear growth (PNG) model in $1+1$ dimension with flat initial condition and no extra constraints. Through the Robinson-Schensted-Knuth (RSK) construction, one obtains the multilayer PNG model, which consists of a stack of non-intersecting lines, the top one being the PNG height. The statistics of the lines is translation invariant and at a fixed position the lines define a point process. We prove that for large times the edge of this point process, suitably scaled, has a limit. This limit is a Pfaffian point process and identical to the one obtained from the edge scaling of Gaussian orthogonal ensemble (GOE) of random matrices. Our results give further insight to the universality structure within the KPZ class of $1+1$ dimensional growth models.
\end{abstract}

\section{Introduction}\label{sect1}
The polynuclear growth (PNG) model is the best-studied growth model from the KPZ class in one spatial dimension. Since in one dimension the dynamical scaling exponent is $z=3/2$, the correlation length increases as $t^{2/3}$ for large growth time $t$. The exponent is universal, but different classes of initial conditions lead to distinct scale invariant statistical properties of the surface in the large $t$ limit. To go beyond the exponents and to determine the exact scaling functions one has to analyze some solvable models. In this paper we consider the PNG model. The surface height at time $t$ is denoted by $x\mapsto h(x,t)$. On the surface new islands of height one are created at random with intensity $\varrho$. The islands spread with unit speed and simply merge upon contact.

For the PNG model the exact scaling function for stationary growth is known, see~\cite{PS02b}, relying on previous results by Baik and Rains~\cite{BR99,BR00}. In the case where initially $h(x,0)=0$ and nucleations are constrained to occur only above a first spreading layer, the surface has typically the shape of a droplet. In this geometry, for fixed but large $t$, the spatial statistics of the surface is well understood. Subtracting the deterministic part, it is proved in~\cite{PS02} that the self-similar shape fluctuations are governed by the Airy process.

In numerical simulations, one starts the growth process mostly with a flat substrate, i.e., $h(x,0)=0$ with no further constraints. Thus it would be of interest to understand the statistics of $x\mapsto h(x,t)$ at large $t$. It is a space translation invariant process and the only available result~\cite{BR99,PS00} is the one-point distribution,
\begin{equation}\label{eq1.2}
\lim_{t\to\infty} \Pb\left(h(0,t) \leq 2t + \xi t^{1/3}\right)=F_1(\xi 2^{2/3}),
\end{equation}
with $F_1$ the GOE Tracy-Widom distribution (here $\varrho=2$). The limiting function $F_1$ is linked to the Gaussian Orthogonal Ensemble (GOE) of random matrix theory as follows. Let us denote by $\lambda_{\max,N}$ the largest eigenvalue of a $N\times N$ GOE matrix, in units where $\E(\lambda_{\max,N})=2N$. Then
\begin{equation}
\lim_{N\to\infty}\Pb\left(\lambda_{\max,N} \leq 2N + \xi N^{1/3}\right) = F_1(\xi),
\end{equation}
see~\cite{TW96}. $F_1$ is given in terms of a Fredholm determinant, compare (\ref{eq2.8b}) below. A plot of $\dx F_1(\xi)/\dx \xi$ in semi-logarithmic scale is available in~\cite{PS00}.

The result (\ref{eq1.2}) leaves open the joint distribution at two space points, even more the full process with respect to $x$. From the general KPZ scaling theory, see~\cite{PS01} for an exposition, a meaningful limit is expected only if the two points are separated by a distance of order $t^{2/3}$. Thus the issue is to determine the limit
\begin{equation}\label{eq1.4}
\lim_{t\to\infty}\Pb\left(h(0,t)\leq 2t + \xi_1 t^{1/3}, h(\tau t^{2/3},t) \leq 2t+\xi_2 t^{1/3}\right)=\,?
\end{equation}
Of course, the marginals are $F_1(\xi_1 2^{2/3})$ and $F_1(\xi_2 2^{2/3})$. But this leaves many choices for the joint distribution.

In the present contribution we will not succeed in removing the question mark in (\ref{eq1.4}). However, we will make a big step towards a well-founded conjecture. The idea to progress in the direction constituting the main body of our paper was set forward by Kurt Johansson in a discussion taking place at the 2003 workshop on growth processes at the Newton Institute, Cambridge (as communicated by Herbert Spohn). In a somewhat rough description, underlying the PNG process there is a line ensemble constructed through the Robinson-Schensted-Knuth (RSK) algorithm. Its top line at time $t$ is the height $x\mapsto h(x,t)$. Thus the question mark in (\ref{eq1.4}) refers to the top line. But instead of (\ref{eq1.4}) we will study, for large $t$, the line ensemble at fixed $x=0$ close to the top line. As in a way already suggested by the Baik and Rains result, under suitable scaling the lines at $x=0$ have indeed the statistics of the top eigenvalues of GOE random matrices. The implications for (\ref{eq1.4}) will be discussed after explaining more precisely our main result.

\section{Main result}\label{sect2}
First we describe the PNG model with flat initial conditions, and secondly recall some random matrix results on GOE eigenvalues, as needed to state the scaling limit.

\subsection{Polynuclear growth (PNG) model and RSK construction}
The polynuclear growth (PNG) model considered here is a $1+1$ dimensional model. One way to view PNG is via a graphical construction involving Poisson points.
Consider a fixed $T>0$ and let $\omega$ be a countable configuration of points in $\R\times[0,T]$. For any compact subset $B$ of $\R\times[0,T]$, denote the number of points of $\omega$ in $B$ by $n(B)(\omega)$. Then
\begin{equation}
\Omega=\{\omega | n(B)(\omega)<\infty,\forall \textrm{ compact }B\subset \R\times[0,T]\}
\end{equation}
is the set of all locally finite point configurations in $\R\times[0,T]$. The Poisson process with intensity $\varrho>0$ in $\R\times[0,T]$ is given by setting the probability $\Pb_T$ such that
\begin{equation}
\Pb_T(\{n(B)=n\})=\frac{(\varrho |B|)^n}{n!}e^{-\varrho |B|}
\end{equation}
for all compact $B\subset\R\times[0,T]$, and the family of random variables $\{n(B_j), j=1,\ldots,m\}$ with $B_i\cap B_j=\emptyset$ for $i\neq j$, is always independent.
In what follows we set $\varrho=2$.

\begin{figure}[t!]
\begin{center}
\psfrag{x}{$x$}
\psfrag{t}{$t$}
\psfrag{t=T}{$t=T$}
\psfrag{h0}[c]{$h=0$}
\psfrag{h1}[c]{$h=1$}
\psfrag{h2}[c]{$h=2$}
\psfrag{h3}[c]{$h=3$}
\includegraphics[width=10cm]{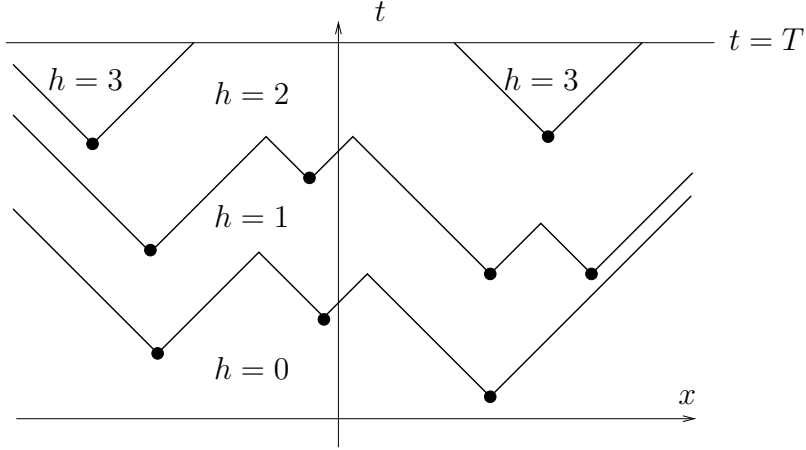}\caption{Graphical construction generating the surface height from the Poisson points.}\label{figRSK2}
\end{center}
\end{figure}

For each $\omega\in\Omega$ we define the height function $h(x,t)(\omega)$, $(x,t)\in \R\times[0,T]$, by the following graphical construction. Because of flat initial conditions, we set $h(x,0)(\omega)=0$ and we call \emph{nucleation events} the points of $\omega$. Each nucleation event generates two lines, with slope $+1$ and $-1$ along its forward light cone. A line ends upon crossing another line. In Figure~\ref{figRSK2} the dots are the nucleation events and the lines follow the forward light cones.
The height $h(x,t)(\omega)$ is then the number of lines crossed along the straight path from $(x,0)$ to $(x,t)$. Since $\omega$ is locally finite, it follows that $x\mapsto h(x,t)(\omega)$, $t\in [0,T]$, is locally bounded and the number of discontinuities is locally finite.

The interpretation of the graphical construction in terms of a growing surface is the following. The surface height at position $x\in\R$ and time $t\geq 0$ is $h(x,t)\in\Z$. The initial condition is $h(x,0)=0$ for all $x\in\R$. For fixed time $t$, consider the height function $x\mapsto h(x,t)$. We say that there is an up-step (of height one) at $x$ if $h(x,t)=\lim_{y\uparrow x}h(y,t)+1$ and a down-step (of height one) at $x$ if $h(x,t)=\lim_{y\downarrow x}h(y,t)+1$. A nucleation event which occurs at position $x$ and time $t$ is a creation of a pair of up- and down-step at $x$ at time $t$. The up-steps move to the left with unit speed and the down-steps to the right with unit speed. When a pair of up- and down-step meet, they simply merge. In Figure~\ref{figRSK2} the dots are the nucleation events, the lines with slope $-1$ (resp.\ $+1$) are the positions of the up-steps (resp.\ down-steps). Other initial conditions and geometries can be treated in a similar fashion. For example, if $h(x,0)$ is not $0$ for all $x$, it is enough to add additional lines starting from the $t=0$ axis with slope $\pm 1$ reflecting the up/down direction of the steps at $t=0$. Another interesting situation is the PNG droplet, where one starts with flat initial conditions and there are no nucleation points outside the forward light cone starting at $(0,0)$.

To study the surface height at time $T$, $x\mapsto h(x,T)$, it is convenient to extend to a multilayer model. This is achieved using the RSK construction. We construct a set of height functions $h_\ell(x,t)(\omega)$, $(x,t)\in \R\times [0,T]$, $\ell \leq 0$ as follows. At $t=0$ we set $h_\ell(x,0)=\ell$ with $\ell=0,-1,\ldots$, $\ell$ denoting the level's height. The first height is defined by $h_{0}(x,t)(\omega)\equiv h(x,t)(\omega)$. The meeting points of the forward light cones generated by the points of $\omega$ are called the \emph{annihilation events} of level $0$. $h_{-1}(x,t)(\omega)$ is constructed as $h_{0}(x,t)(\omega)$ but the nucleation events for level $-1$ are the annihilation events of level $0$ and $h_{-1}(x,t)(\omega)+1$ equals the number of lines for level $-1$ crossed from $(x,0)$ to $(x,t)$.
\begin{figure}[t!]
\begin{center}
\psfrag{x}{$x$}
\psfrag{t}{$t$}
\psfrag{t=T}{$t=T$}
\psfrag{1}{$\ell=-1$}
\psfrag{2}{$\ell=-2$}
\psfrag{3}{$\ell=-3$}
\psfrag{4}{$\ell=-4$}
\includegraphics[width=10cm]{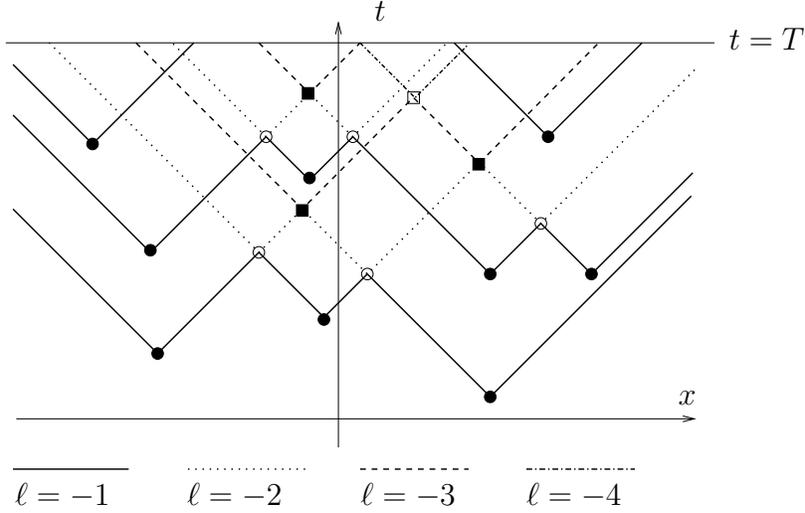}\caption{RSK construction up to time $t=T$.}\label{figRSK3a}
\end{center}
\end{figure}
In Figure~\ref{figRSK3a} the nucleation events of level $-1$ are the empty dots, whose forward light cones are the dotted lines. Setting the annihilation events of level $j$ as the nucleation events for level $j-1$, the set of height functions $h_\ell(x,t)(\omega)$ is defined for all $\ell \leq 0$. The line ensemble for $t=T$, i.e., $\{h_\ell(x,T), \ell \leq 0\}$ is represented in Figure~\ref{figRSK3b}.
\begin{figure}[t!]
\begin{center}
\psfrag{x}{$x$}
\psfrag{j}{$j$}
\psfrag{h1}[r]{$h_{0}$}
\psfrag{h2}[r]{$h_{-1}$}
\psfrag{h3}[r]{$h_{-2}$}
\psfrag{h4}[r]{$h_{-3}$}
\psfrag{h5}[r]{$h_{-4}$}
\includegraphics[width=10cm]{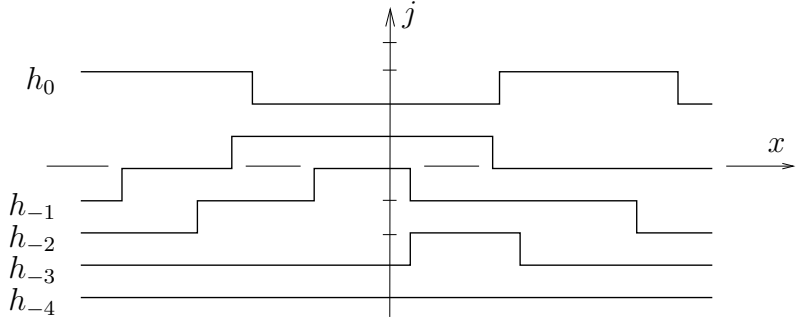}\caption{Line ensemble for $t=T$ for the point configuration of Figure~\ref{figRSK3a}.}\label{figRSK3b}
\end{center}
\end{figure}

The point process which describes this line ensemble at $x=0$ is denoted by $\zeta_T^{\rm flat}$ and given by
\begin{equation}\label{eq2.1b}
\zeta_T^{\rm flat}(j)=\left\{\begin{array}{ll}1&\textrm{ if a line passes at }(0,j),\\0&\textrm{ if no line passes at }(0,j).\end{array}\right.
\end{equation}
From the Baik and Rains result we know that the largest $j$ such that $\zeta_T^{\rm flat}(j)\neq 0$ is located near $2T$ and fluctuates on a $T^{1/3}$ scale. The edge rescaled point process is defined as follows. For any smooth test function $f$ of compact support
\begin{equation}\label{eq2.2b}
\eta_T^{\rm flat}(f)=\sum_{j\in\Z}\zeta_T^{\rm flat}(j) f\big((j-2T)/(T^{1/3}2^{-2/3})\big),
\end{equation}
the factor $2^{-2/3}$ is the same as in (\ref{eq1.2}). Notice that in (\ref{eq2.2b}) there is no prefactor to the sum. The reason is that close to $2T$, the points of $\zeta_T^{\rm flat}$ are order $T^{1/3}$ apart and $\eta_T^{\rm flat}$ remains a point process in the limit $T\to\infty$. $\eta_T^{\rm flat}$ has a last particle, i.e., $\eta_T^{\rm flat}(\xi)=0$ for all $\xi$ large enough, and even in the $T\to\infty$ limit has a finite density which increases as $\sqrt{-\xi}$ as $\xi\to -\infty$. Consequently the sum in (\ref{eq2.2b}) is effectively finite.

\subsection{Random matrices}\label{RM}
The Gaussian Orthogonal Ensemble (GOE) is the set of $N\times N$ real symmetric matrices distributed according to the probability measure $Z_N^{-1}\exp[-\Tr{M^2}/2N]\dx M$, with $\dx M=\prod_{1\leq i \leq j \leq N}\dx M_{i,j}$. The eigenvalues are then distributed with density
\begin{equation}
\frac{1}{Z'_N}\prod_{1\leq i < j \leq N}|\lambda_i-\lambda_j| \prod_{i=1}^N\exp[-\lambda_i^2/2N].
\end{equation}
Let us denote by $\zeta_N^{\rm GOE}$ the point process of GOE eigenvalues, i.e., $\zeta_N^{\rm GOE}(x)=\sum_{j=1}^N\delta(x-\lambda_j)$. At the edge of the spectrum, $2N$, the eigenvalues are order $N^{1/3}$ apart. The edge rescaled point process is then given by
\begin{equation}\label{eq1.7}
\eta_N^{\rm GOE}(\xi)= N^{1/3} \zeta_N^{\rm GOE}(2N+\xi N^{1/3}),
\end{equation}
and for $f$ a test function of compact support,
\begin{equation}\label{eq1.8}
\eta_N^{\rm GOE}(f)= \sum_{j=1}^N f\big((\lambda_j-2N)/N^{1/3}\big) =\int_{\R}\dx\xi f(\xi) \eta_N^{\rm GOE}(\xi).
\end{equation}
We denote by $\eta^{\rm GOE}$ the limit of $\eta_N^{\rm GOE}$ as $N\to\infty$.

The limit point process $\eta^{\rm GOE}$ is characterized by its correlation functions as follows.
Let us denote by $\rho^{(n)}_{\rm GOE}(\xi_1,\ldots,\xi_n)$ the $n$-point correlation functions of $\eta^{\rm GOE}$, i.e., the joint density of having eigenvalues at $\xi_1,\ldots,\xi_n$. Then
\begin{equation}\label{eq2.6b}
\rho^{(n)}_{\rm GOE}(\xi_1,\ldots,\xi_n)=\Pf[G^{\rm GOE}(\xi_i,\xi_j)]_{i,j=1,\ldots,n}
\end{equation}
where $\Pf$ is the Pfaffian and $G^{\rm GOE}$ is a $2\times 2$ matrix kernel with elements
\begin{eqnarray}\label{eqGOEkernel}
G_{1,1}^{\rm GOE}(\xi_1,\xi_2)
&=&\int_0^{\infty} \dx\lambda \Ai(\xi_1+\lambda)\Aip(\xi_2+\lambda)-(\xi_1\leftrightarrow \xi_2), \\
G_{1,2}^{\rm GOE}(\xi_1,\xi_2) 
&=&\int_0^{\infty} \dx\lambda \Ai(\xi_1+\lambda)\Ai(\xi_2+\lambda)+\frac12 \Ai(\xi_1) \int_0^{\infty}\dx\lambda \Ai(\xi_2-\lambda), \nonumber\\
G_{2,1}^{\rm GOE}(\xi_1,\xi_2) 
&=&-G_{1,2}^{\rm GOE}(\xi_2,\xi_1) \nonumber\\
G_{2,2}^{\rm GOE}(\xi_1,\xi_2) 
&=&\frac14 \int_0^\infty\dx\lambda\int_\lambda^\infty\dx\mu \Ai(\xi_2-\mu)\Ai(\xi_1-\lambda)-(\xi_1\leftrightarrow \xi_2), \nonumber
\end{eqnarray}
and $\Ai$ is the Airy function~\cite{AS84}. The notation $(\xi_1\leftrightarrow \xi_2)$ means that the previous term is repeated with $\xi_1$ and $\xi_2$ interchanged. For an antisymmetric matrix $A$, $\Pf(A)=\sqrt{\Det(A)}$, see (\ref{eq3.1}) for the definition of the Pfaffian. The GOE kernel was studied in~\cite{TW96}. It is not uniquely defined, for example the one reported in~\cite{FNH99,SI03} differs slightly from the one written here, but they are equivalent because they yield the same point process. The point process $\eta^{\rm GOE}$ is uniquely determined by its correlation functions~\cite{Sos00}.

Finally let us remark that $F_1$ can be written in terms of a Fredholm determinant
\begin{eqnarray}\label{eq2.8b}
F_1(\xi)&=&\lim_{N\to\infty}\E_N\bigg(\prod_{j=1}^N (1-\Id_{(\xi,\infty)}((\lambda_j-2N)/N^{1/3}))\bigg)\nonumber \\
&=&\sqrt{\Det(\Id- J^{-1} G^{\rm GOE})},
\end{eqnarray}
where $J$ is the matrix kernel $J=\left(\begin{array}{cc}0&1\\ -1 & 0 \end{array}\right)$. The determinant in (\ref{eq2.8b}) is the Fredholm determinant of the kernel $K=J^{-1} G^{\rm GOE}$ on the measure space $((\xi,\infty)\times \{1,2\},\dx x \times \nu)$ with $\dx x$ the Lebesgue measure and $\nu$ the counting measure on $\{1,2\}$, i.e., 
\begin{equation}\label{eqSeries}
\Det(\Id-K)=\sum_{k=0}^\infty \frac{(-1)^n}{n!}\int_{(\xi,\infty)^n}\hspace{-0.3cm}\dx\xi_1\cdots\dx\xi_n
\hspace{-0.3cm}\sum_{i_1,\ldots,i_n\in\{1,2\}} \hspace{-0.3cm}\Det\left[K_{i_k,i_l}(\xi_k,\xi_l)\right]_{k,l=1,\ldots,n}.
\end{equation}

Remark: One can also consider instead of $\Det(\Id-K)$ the determinant $\Det(\Id-\hat K)$ with $\hat K$ the operator with kernel $K$. $\hat K$ is not trace-class on $L^2((\xi,\infty),\dx x)\oplus L^2((\xi,\infty),\dx x)$ because it is not even Hilbert-Schmidt. Nevertheless it is possible to make sense of it as follows. $\hat K$ is Hilbert-Schmidt in the space $L^2((\xi,\infty),\theta^{-1} \dx x)\oplus L^2((\xi,\infty),\theta\dx x)$ where $\theta$ is any positive weight function growing at most polynomially at $x\to\infty$ and satisfying $\theta^{-1}\in L^1((\xi,\infty),\dx x)$. Moreover the sum of the diagonal terms, $\tilde \Tr(\hat K)$, is absolutely integrable. Then the modified Fredholm determinant, which has also the series development (\ref{eqSeries}), is defined by $\Det(\Id-\hat K)=e^{-\tilde\Tr(\hat K)} \Det_2(\Id-\hat K)$ with $\Det_2$ the regularized determinant~\cite{GK69}. This functional analysis point of view is used by Tracy and Widom~\cite{TW04} to show the convergence of the modified Fredholm determinants in the $N\to\infty$ limit.

\subsection{Scaling limit}
As our main result we prove that the point process $\eta_T^{\rm flat}$ converges weakly to the point process $\eta^{\rm GOE}$ as $T\to\infty$.
\begin{theorem}\label{thmMain} For any $m\in\N$ and smooth test functions of compact support $f_1,\ldots,f_m$,
\begin{equation}\label{eqthmMain}
\lim_{T\to\infty}\E_T\bigg(\prod_{k=1}^m\eta_T^{\rm flat}(f_k)\bigg) =\E\bigg(\prod_{k=1}^m\eta^{\rm GOE}(f_k)\bigg).
\end{equation}
\end{theorem}
$\E_T$ refers to expectation with respect to the Poisson process measure $\Pb_T$. The expected value on the r.h.s.\ of (\ref{eqthmMain}) is computed via the correlation functions (\ref{eq2.6b}).

As announced in the Introduction, the result of Theorem~\ref{thmMain} is a first step towards a conjecture on the self-similar statistics of the PNG with flat initial conditions. The starting observation is that, as for the PNG, also to random matrices one can introduce a line ensemble in a natural way. Let $M$ be a $N\times N$ random matrix in the GOE, resp.\ GUE, ensemble. As noticed by Dyson~\cite{Dys62} when the coefficients of $M$ are independent Ornstein-Uhlenbeck processes, then the eigenvalues $\lambda_j(t)$ of $M=M(t)$ satisfy the set of stochastic differential equations
\begin{equation}\label{eq2.9}
\dx\lambda_j(t) =
\bigg(-\frac{1}{2N}\lambda_j(t) + \frac{\beta}{2} \sum^N_{\begin{subarray}{l}i=1,\\
i\neq j\end{subarray}}
\frac{1}{\lambda_j(t)- \lambda_i(t)} \bigg)\dx t + \dx b_j(t)\,, \quad
j=1,...,N,
\end{equation}
with $\{b_j(t),\, j=1,...,N\}$ a collection of $N$ independent standard Brownian motions, $\beta=1$ for GOE and $\beta=2$ for GUE. We refer to the \emph{stationary process} of (\ref{eq2.9}) as \emph{Dyson's Brownian motion}. Note that for $\beta\geq 1$ there is no crossing of the eigenvalues, as proved by Rogers and Shi~\cite{RS93}.

Let us denote by $\zeta_N^{\mathrm{GUE}}$ the point process of GUE eigenvalues, i.e., $\zeta_N^{\mathrm{GUE}}(x,t)=\sum_{j=1}^N\delta(x-\lambda_j(t))$ where $\lambda_j(t)$ are the GUE eigenvalues at time $t$. At the edge of the spectrum the eigenvalues are order $N^{1/3}$ apart and the space-time edge rescaled point process is given by
\begin{equation}\label{eq2.12b}
\eta_N^{\mathrm{GUE}}(\xi,\tau)= N^{1/3} \zeta_N^{\mathrm{GUE}}(2N+\xi N^{1/3},2 \tau N^{2/3}).
\end{equation}
Its limit as $N\to\infty$ is denoted by $\eta^{\mathrm{GUE}}(\xi,\tau)$. $\eta^{\mathrm{GUE}}$ is a determinantal point process and its space-time kernel is the extended Airy kernel. The top line of $\eta^{\mathrm{GUE}}$ is given by the Airy process, denoted by $\mathcal{A}(t)$, which appears in~\cite{PS02,Jo03} with more detailed properties investigated in~\cite{AvM03,TW03,TW03b,Wid03}. The height statistics $x\mapsto h(x,t)$ for the PNG \emph{droplet} is linked to the Airy process by
\begin{equation}
\lim_{T\to\infty} T^{-1/3}\Big(h(\tau T^{2/3},T)-2\sqrt{T^2-(\tau T^{2/3})^2}\Big) = \mathcal{A}(\tau),
\end{equation}
where the term subtracted from $h$ is the asymptotic shape of the droplet~\cite{PS02}. To obtain this result, Pr\"ahofer and Spohn consider the line ensemble obtained by RSK and define a point process like (\ref{eq2.1b}) but extended to space-time. It is a determinantal point process and in the edge scaling it converges, as $T\to\infty$, to the point process associated with the extended Airy kernel. Thus they prove not only that the top line converges to the Airy process, but also that the top lines converge to the top lines of Dyson's Brownian motion with $\beta=2$.

One can extend $\eta_N^{\rm GOE}$ of (\ref{eq1.7}) to space-time as in (\ref{eq2.12b}). The conjecture is that, under edge scaling, the process $x\mapsto h(x,T)$ for flat PNG is in distribution identical to the largest eigenvalue of Dyson's Brownian motion with $\beta=1$. The result of Theorem~\ref{thmMain} makes this conjecture more plausible. In fact we now know that, not only $h(0,T)$ in the limit $T\to\infty$ and properly rescaled  is GOE Tracy-Widom distributed, but also that the complete point process $\eta_T^{\rm flat}$ converges to the edge scaling of Dyson's Brownian motion with $\beta=1$ for fixed time. For $\beta=1$ Dyson's Brownian motion one expect that under edge scaling the full stochastic process has a limit. More explicitly, one focuses at the space-time point $(2N,0)$, rescales space by a factor $N^{1/3}$, time by $N^{2/3}$, and expects that the statistics of the lines has a limit for $N\to\infty$. It could be that this limit is again Pfaffian with suitably extended kernel. But even for $\beta=1$ Dyson's Brownian motion this structure has not been unravelled.

The rest of the paper is organized as follows. In Section~\ref{sect3} we explain the line ensemble which will be used to obtain our result. It differs from the one of Figure~\ref{figRSK3b}. The end points of the line ensemble gives a point process, whose correlation functions are obtained in Section~\ref{sect4}. They are given in term of a $2\times 2$ matrix kernel which is computed for fixed $T$ in Section~\ref{sect5}. Section~\ref{sect6} is devoted to the edge scaling of the kernel and its asymptotics. Finally Section~\ref{sect7} contains the proof of Theorem~\ref{thmMain}. 

\section{Line ensemble}\label{sect3}

\subsection{Line ensemble for the $\boxbslash$ symmetry}
\begin{figure}[t!]
\begin{center}
\psfrag{x}[c]{$x$}
\psfrag{t}[r]{$t$}
\psfrag{s}[l]{$s$}
\psfrag{j}[l]{$j$}
\psfrag{g0}[l]{$\gamma(0)$}
\psfrag{gs}[l]{$\gamma(s)$}
\psfrag{gT}[l]{$\gamma(T\sqrt{2})$}
\psfrag{H0}[l]{$H_0$}
\psfrag{H1}[l]{$H_{-1}$}
\psfrag{H2}[l]{$H_{-2}$}
\includegraphics[width=\textwidth]{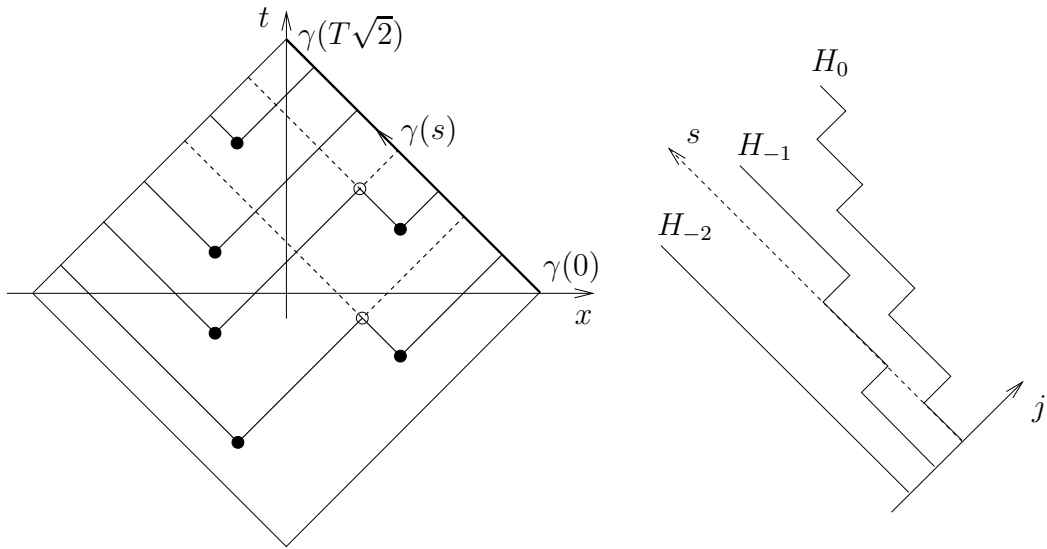}\caption{A configuration with three Poisson points in the triangle $\triangle_+$ and their symmetric images with respect to the $t=0$ axis. The path $\gamma$ is the bold line. On the right we draw the top lines of the line ensemble associated to $\gamma$, $\{H_j(s),j\leq 0,s\in[0,T\sqrt{2}]\}$.} \label{figSymmetric}
\end{center}
\end{figure}

The line ensemble for flat PNG generated by RSK at time $t=T$ is not easy to analyze because there are non-local constraints on the line configurations.
Instead, we start considering the point process $\zeta_T^{\rm flat}$. First remark that this point process depends only on the points in the triangle
$\triangle_+=\{(x,t)\in\R\times\R_+ | t\in [0,T], |x|\leq T-t\}$. We then consider the Poisson points only in $\triangle_+$ and add their symmetric images with respect to the $t=0$ axis, which are in $\triangle_-=\{(x,t)\in\R\times\R_- | (x,-t)\in\triangle_+\}$. We denote by $\zeta_T^{\textrm{sym}}$ the point process at $(0,T)$ obtained by RSK construction using the Poisson points and their symmetric images, see Figure~\ref{figSymmetric}. 
To study $\zeta_T^{\textrm{sym}}$ we consider a \emph{different line ensemble}. Let us consider the path in space-time defined by $\gamma(s)=(T-s/\sqrt{2},s/\sqrt{2})$, $s\in[0,T\sqrt{2}]$, and construct the line ensemble $\{H_j(s),j\leq 0, s\in[0,T\sqrt{2}]\}$, as follows. The initial conditions are $H_j(0)=j$ since the height at $t=0$ is zero everywhere. Every times that $\gamma$ crosses a RSK line corresponding to a nucleation event of level $j$,  $H_j$ has an up-jump. Then the point process $\zeta_T^{\textrm{sym}}$ is given by the points $\{H_j(T\sqrt{2}),j\leq 0\}$. In Proposition~\ref{prop1} we show that $\zeta_T^{\mathrm{flat}}$ can be recovered by $\zeta_T^{\textrm{sym}}$, in fact we prove that $h_j(0,T)=\frac12 (H_j(T\sqrt{2})+j).$

Next we have to determine the allowed line configurations and their distribution induced by the Poisson points. This is obtained as follows. We prove that the \emph{particle-hole} transformation on the line ensemble $\{H_j(s),j\leq 0,s\in[0,T\sqrt{2}]\}$ is equivalent to a particular change of symmetry in the position of the nucleation events, and we connect with the half-droplet PNG problem studied by Sasamoto and Imamura~\cite{SI03}.

\subsubsection*{Young tableaux}
\begin{figure}[t!]
\begin{center}
\psfrag{S}[c]{$S(\sigma)$}
\psfrag{T}[c]{$T(\sigma)$}
\psfrag{S2}[c]{$S(\tilde\sigma)$}
\psfrag{T2}[c]{$T(\tilde\sigma)$}
\includegraphics[height=6cm]{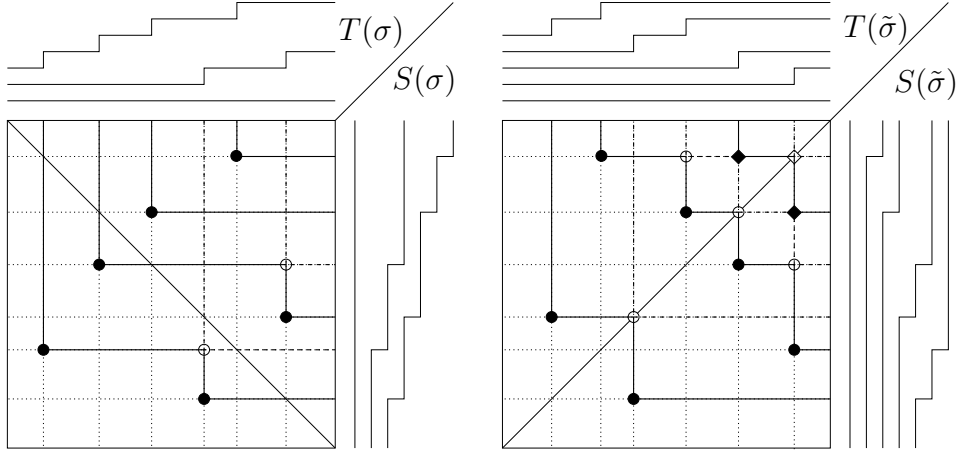}\caption{Line ensembles for $\sigma=(2\,4\,5\,1\,6\,3)$ and $\tilde\sigma=(3\,6\,1\,5\,4\,2)$. For the even levels, $\ell=0,-2$, we use the solid lines and for the odd levels, $\ell=-1,-3$, the dashed lines. The line ensemble of $S(\sigma)$ corresponds to the line ensemble $\{H_j(s),j\leq 0,s\in[0,T\sqrt{2}]\}$ of Figure~\ref{figSymmetric}.} \label{figYoung}
\end{center}
\end{figure}

Let $\sigma=(\sigma(1),\ldots,\sigma(2N))$ be a permutation of $\{1,\ldots,2N\}$ which indicate the order in which the Poisson points are placed in the diamond $\triangle_+\cup\triangle_-$. More precisely, let $(x_i,t_i)$ be the positions of the points with the index $i=1,\ldots,N$ such that $t_i+x_i$ is increasing with $i$, and $\sigma$ is the permutation such that $t_{\sigma(i)}-x_{\sigma(i)}$ is increasing in $i$ too. Let us construct the line ensembles along the paths $(T,0)\to(0,T)$ and $(-T,0)\to(0,T)$. The relative position of the steps on the line ensembles are encoded in the Young tableaux $S(\sigma)$ and $T(\sigma)$ constructed using Schensted's algorithm. If the $k^{\rm th}$ step occurs in line $H_j$, then in the Young tableau there is a $k$ in row $j$, see Figure~\ref{figYoung}.

In our case the points are symmetric with respect to the axis $t=0$ and we refer to it as the symmetry $\boxbslash$. In the case studied in~\cite{SI03}, the points are symmetric with respect to the axis $x=0$ and we call it the symmetry $\boxslash$. Consider a configuration of points with symmetry $\boxbslash$ and let $\sigma$ be the corresponding permutation. The RSK construction leads to the line ensembles of $T(\sigma)$ and $S(\sigma)$ as shown in the left part of Figure~\ref{figYoung}. If we apply the axis symmetry with respect to $x+t=0$, then we obtain a configuration of points shown in the right part of Figure~\ref{figYoung}. The points have now the symmetry $\boxslash$ and the corresponding permutation $\tilde\sigma$ is obtained simply by reversing the order of $\sigma$, that is, if $\sigma=(\sigma(1),\ldots,\sigma(2N))$ then $\tilde\sigma(j)=\sigma(2N+1-j)$. By Schensted's theorem~\cite{Sch61},
\begin{equation}
S(\tilde\sigma)=S(\sigma)^t,.
\end{equation}
Moreover, the positions of the steps in the line ensembles of $S(\sigma)$ and $S(\tilde\sigma)$ occurs at the same positions, but of course in different line levels. Figure~\ref{figYoung} shows an example with $\sigma=(2\,4\,5\,1\,6\,3)$, for which the Young tableaux are
\begin{center}
\begin{tabular}{cc}
$\begin{array}{c}
S(\sigma)=\left(\begin{array}{cccc} 1&3&5&6\\ 2 &4 \end{array}\right),\\[10pt] T(\sigma)=\left(\begin{array}{cccc} 1&2&3&5\\ 4&6 \end{array}\right),
\end{array}$
&
$ S(\tilde\sigma)=T(\tilde\sigma)=\left(\begin{array}{cc} 1&2\\ 3&4\\5\\6 \end{array}\right).$
\end{tabular}
\end{center}

\subsubsection*{Particle-hole transformation}
\begin{figure}[t!]
\begin{center}
\psfrag{j}[r]{$j$}
\psfrag{0}[r]{$0$}
\psfrag{1}[r]{$1$}
\includegraphics[height=4cm,width=8cm]{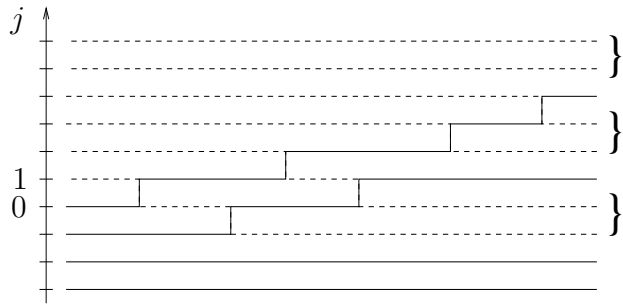}\caption{Particle (solid) and hole (dashed) line ensembles for the example of Figure~\ref{figYoung}. The particle line ensemble is the one associated with $S(\sigma)$, and the hole line ensemble is the one of $S(\tilde\sigma)$ reflected with respect to the line $j=1/2$. The pairing rule is shown by the brackets.}\label{figLines2}
\end{center}
\end{figure}

At the level of line ensemble we can apply the particle-hole transformation, which means that a configuration of lines is replaced by the one with jumps at the same positions and the horizontal lines occupy the previous empty spaces, as shown in Figure~\ref{figLines2}. 
Let us start with the line ensemble corresponding to $S(\sigma)$, then the Young tableau for the \emph{hole line ensemble} is given by $S(\sigma)^t$. In fact, the information encoded in $S(\sigma)$ tell us that the $j^{\textrm{th}}$ particle has jumps at (relative) position $S(\sigma)_{j,k}$, $k\geq 1$, for $j \geq 1$. On the other hand, the $j^{\textrm{th}}$ hole has jumps where the particles have their $j^{\textrm{th}}$ jump. Therefore the particle-hole transformation is equivalent to the symmetry transformation $\boxbslash\to\boxslash$.

\subsubsection*{Allowed line configurations and measure}
Sasamoto and Imamura~\cite{SI03} study the half-droplet geometry for PNG, where nucleation events occurs symmetrically with respect to $x=0$, i.e., with the $\boxslash$ symmetry. In particular, they prove that the point process at $x=0$ converges to the point process of eigenvalues of the Gaussian Symplectic Ensemble (GSE). Its correlation functions have the same Pfaffian structure as GOE but with a different kernel. In a way the line ensemble they study is the hole line ensemble described above, thus their edge scaling focuses at the top holes, i.e., in the region where the lowest particles are excited. Notice that the change of focus between particles and holes changes the statistics from GSE to GOE. This differs from the case of the PNG droplet~\cite{PS02} where for both holes and particles the edge statistics is GUE. Although the result of~\cite{SI03} cannot be applied directly to our symmetry, some properties derived there will be of use.

From~\cite{SI03} we know that for the symmetry $\boxslash$ a hole line configuration is allowed if: a) the lines do not intersect, b) have only down-jumps, c) they satisfy the \emph{pairing rule}: $H^{\rm hole}_{2j}(T\sqrt{2})=H^{\rm hole}_{2j-1}(T\sqrt{2})$ for all $j\geq 1$. This implies that for the symmetry $\boxbslash$ a line configuration $\{H_j\}$ is allowed if: a) the lines do not intersect, b) have only up-jumps, c) $H_j(T\sqrt{2})-H_j(0)$ is \emph{even} for each $j \leq 0$.
Moreover, there is a one-to-one correspondence between allowed configurations and nucleation events. The probability measure for the line ensemble turns out to have a simple structure. Consider Poisson points with intensity $\varrho$ and symmetry $\boxbslash$. Each Poisson point $(x,t)\in \triangle_+$ has a probability $\varrho \, \dx x\dx t$ of being in $[x,x+\dx x]\times[t,t+\dx t]$. In the corresponding line ensemble this weight is carried by two jumps, therefore the measure induced by the points on a line configuration $\{H_j\}$ is given by $\sqrt{\varrho}^{\,\#\textrm{jumps in }\{H_j\}}$ times the uniform measure.

\subsection{Flat PNG and line ensemble for $\boxbslash$ symmetry}
The correspondence between the point process $\zeta_T^{\rm flat}$ and $\zeta_T^{\rm sym}$ is as follows.
Let us consider a permutation $\sigma$ with Young tableau $S(\sigma)$ of shape $(\lambda_1,\lambda_2,\ldots,\lambda_m)$. Let, for $k\leq m$, $a_k(\sigma)$ be the length of the longest subsequence consisting of $k$ disjoint increasing subsequences.
\begin{theorem}[Greene~\cite{Gre74}] For all $k=1,\ldots,m$, 
\begin{equation}
a_k(\sigma)=\lambda_1+\cdots+\lambda_k.
\end{equation}
\end{theorem}

The geometric interpretation is the following. Let $\sigma$ be the permutation which corresponds to some configuration of Poisson points in $\triangle_+\cup\triangle_-$. Then $a_k$ is the maximal sum of the lengths of $k$ non-intersecting (without common points) directed polymers from $(0,-T)$ to $(0,T)$. 

\begin{proposition}\label{prop1} Let $\pi$ be a Poisson point configuration in $\triangle_+$ and let the corresponding Young tableau $S(\pi)$ have shape $(\lambda_1,\lambda_2,\ldots,\lambda_m)$. Let $\tilde \pi$ be the configuration of points on $\triangle_+\cup\triangle_-$ with symmetry $\boxbslash$ which is identical to $\pi$ in $\triangle_+$. Then $S(\tilde \pi)$ has shape $(\tilde \lambda_1,\tilde \lambda_2,\ldots,\tilde \lambda_m)=(2\lambda_1,2\lambda_2,\ldots,2\lambda_m)$.
\end{proposition}
\begin{proof}
To prove the proposition is enough to prove that $a_k(\tilde \pi)=2 a_k(\pi)$ for $k=1,\ldots,m$.\\
i) $a_k(\tilde \pi) \geq 2 a_k(\pi)$: it is obvious since we can choose the $k$ directed polymers on $\tilde \pi$ by completing the ones on $\pi$ by symmetry.\\
ii) $a_k(\tilde \pi) \leq 2 a_k(\pi)$: assume it to be false. Then there exists $k$ directed polymers in $\triangle_+$ and $k$ in $\triangle_-$ such that the total length is strictly greater than $2 a_k(\pi)$. This implies that at least one (by symmetry both) of the sets of $k$ directed polymers has total length strictly greater that $a_k(\pi)$. But this is in contradiction with the definition of $a_k(\pi)$, therefore $a_k(\tilde \pi) \leq 2 a_k(\pi)$.
\end{proof}
Since $\lambda_{1-j}=h_j(0,T)-j$ and $\tilde \lambda_{1-j} = H_j(T\sqrt{2})-j$, it follows from this proposition that
\begin{equation}
h_j(0,T)=\tfrac12(H_j(T\sqrt{2})+j)
\end{equation}
for all $j\leq 0$.

\section{Correlation functions}\label{sect4}
Non-intersecting lines can be viewed as trajectories of fermions in discrete space $\Z$ and continuous time $[0,T\sqrt{2}]$. Let us start with a finite number of fermions, $2N$, which implies that only the information in the first $2N$ levels in the RSK construction is retained. For any configuration, the number of non perfectly flat lines, is obviously bounded by the number of Poisson points in $\triangle_+$. On the other hand for fixed $T$, the probability of having a number of Poisson points greater than $2N$ decreases exponentially fast for $N$ large. First we derive an exact formula for the $n$-point correlation function for finite $N$, and then take the limit $N\to\infty$ so that, for any fixed $T$, each line configuration contains all the information of the Poisson points. Finally we consider the asymptotic for large $T$.

\subsection{Correlation functions and Pfaffians}
The correlation functions of the point process $\zeta_T^{\rm sym}$ turn out to be expressed as Pfaffians. Therefore we first review the Pfaffian ensemble. Let $A=[A_{i,j}]_{i,j=1,\ldots,2N}$ be an \emph{antisymmetric} matrix, then its Pfaffian is defined by
\begin{equation}\label{eq3.1}
\Pf(A)=\sum_{\begin{subarray}{c}\sigma\in S_{2N}\\ \sigma_{2i-1}<\sigma_{2i}\end{subarray}} (-1)^{|\sigma|} \prod_{i=1}^N A_{\sigma_{2i-1},\sigma_{2i}},
\end{equation}
where $S_{2N}$ is the set of all permutations of $\{1,\ldots,2N\}$.
Notice that the Pfaffian depends only on the upper triangular part of $A$. For an antisymmetric matrix the identity $\Pf(A)^2=\Det(A)$ holds.

The Pfaffian ensemble is introduced in~\cite{Ra00}, see also~\cite{Sos03}. Let $(X,\mu)$ be a measure space, $f_1,\ldots,f_{2N}$ complex-valued functions on $X$, $\e(x,y)$ be an \emph{antisymmetric kernel}, and define by
\begin{equation}\label{eq3.2}
p(x_1,\ldots,x_{2N})=\frac{1}{Z_{2N}}\Det[f_j(x_k)]_{j,k=1,\ldots,2N} \Pf[\e(x_j,x_k)]_{j,k=1,\ldots,2N}
\end{equation}
the density of a $2N$-dimensional probability distribution on $X^{2N}$ with respect to $\mu^{\otimes 2N}$, the product measure generated by $\mu$.
The normalization constant is given by
\begin{equation}
Z_{2N}=\int_{X^{2N}}\dx^{2N}\mu \Det[f_j(x_k)]_{j,k=1,\ldots,2N} \Pf[\e(x_j,x_k)]_{j,k=1,\ldots,2N}=(2N)! \Pf[M]
\end{equation}
where the matrix $M=[M_{i,j}]_{i,j=1,\ldots,2N}$ is defined by
\begin{equation}
M_{i,j}=\int_{X^2}f_i(x)\e(x,y)f_j(y)\dx\mu(x)\dx\mu(y).
\end{equation}

The $n$-point correlation functions $\rho^{(n)}(x_1,\ldots,x_n)$ of a point process with measure (\ref{eq3.2}) are given by Pfaffians
\begin{equation}
\rho^{(n)}(x_1,\ldots,x_n)=\Pf[K(x_i,x_j)]_{i,j=1,\ldots,n}
\end{equation}
where $K(x,y)$ is the antisymmetric kernel
\begin{equation}
K(x,y)=\left(\begin{array}{cc}K_{1,1}(x,y) & K_{1,2}(x,y)\\ K_{2,1}(x,y) & K_{2,2}(x,y)\end{array}\right)
\end{equation}
with
\begin{equation}
\begin{array}{rcl}
K_{1,1}(x,y)&=&\sum_{i,j=1}^{2N} f_i(x) M_{j,i}^{-1} f_j(y)\\[6pt]
K_{1,2}(x,y)&=&\sum_{i,j=1}^{2N} f_i(x) M_{j,i}^{-1} (\e f_j)(y)\\[6pt]
K_{2,1}(x,y)&=&\sum_{i,j=1}^{2N} (\e f_i)(x) M_{j,i}^{-1} f_j(y)\\[6pt]
K_{2,2}(x,y)&=&-\e(x,y)+\sum_{i,j=1}^{2N} (\e f_i)(x) M_{j,i}^{-1} (\e f_j)(y)
\end{array}
\end{equation}
provided that $M$ is invertible, and $(\e f_i)(x)= \int_X \e(x,y) f_i(y)\dx\mu(y)$. Note the order of indices in $M_{j,i}^{-1}$.
 
\subsection{Linear statistics}
Let $a^*_j$ and $a_j$, $j\in\Z$, be the creation and annihilation operator for the fermions and $\ket{\emptyset}$ be the state without fermions. The initial state is then given by
\begin{equation}
\ket{\Omega_{\rm in}} = \prod_{j=-2N+1}^{0} a^*_j \ket{\emptyset},
\end{equation}
and the final state is 
\begin{equation}\label{eq4.2b}
\ket{\Omega_{\rm fin}} = \sum_{{\bf n}\in C_N}\prod_{j=-2N+1}^{0} a^*_{j+2n_j} \ket{\emptyset}
\end{equation}
where $C_N=\{\{n_{0},\ldots,n_{-2N+1}\}|\, n_{j} \geq n_{j-1}, n_{j}\geq 0\}$.
Let us define the up-jump operator as
\begin{equation}
\alpha_1=\sum_{k\in\Z}a^*_{k+1}a_k,
\end{equation}
which when applied on $\ket{\Omega_{\rm in}}$ is actually a finite sum. Then the evolution from the initial state $(t=0)$ to the final one $(t=T\sqrt{2})$ is given by the transfer operator
\begin{equation}
\exp(\tilde T\alpha_1),\quad \tilde T=\sqrt{2\varrho}T=2T.
\end{equation}
The linear statistics, i.e., for a bounded function $g:\Z\to\R$, is
\begin{equation}
\E_{N,T}\bigg(\prod_{j=-2N+1}^{0}(1-g(x_j^{\rm fin}))\bigg) = \frac{\bra{\Omega_{\rm fin}} \prod_{y\in\Z}(1-g(y)a^*_y a_y) e^{\tilde T\alpha_1} \ket{\Omega_{\rm in}}}{\bra{\Omega_{\rm fin}} \prod_{y\in\Z}e^{\tilde T\alpha_1}\ket{\Omega_{\rm in}}}
\end{equation} where the $x_j^{\rm fin}$, $j\in\{-2N+1,\ldots,0\}$ are the position of the fermions at time $T\sqrt{2}$.
Let us denote by $\rho^{(n)}(x_1,\ldots,x_n)$ the $n$-point correlation function of $\zeta_T^{\rm sym}$. Then
\begin{equation}\label{eqFP}
\E_{N,T}\bigg(\prod_{j=-2N+1}^{0}(1-g(x_j^{\rm fin}))\bigg) = \sum_{n\geq 0} \frac{(-1)^n}{n!}
\sum_{x_1,\ldots,x_n\in\Z} \rho^{(n)}(x_1,\ldots,x_n)\prod_{j=1}^n g(x_j).
\end{equation}
For finite $N$, $\rho^{(n)}=0$ for $n>2N$. 

\begin{proposition}\label{PropKernel}
Let us define the matrix $\Phi$ with entries
\begin{equation}\label{eq4.5b}
\Phi_{x,i}= \frac{1}{(x-i)!} \tilde T^{x-i} \Theta(x-i),
\end{equation}
with $\Theta$ the Heaviside function, the antisymmetric matrices $S$ and $A$
\begin{equation}
S_{x,y}=\frac{1+\sgn(x-y) (-1)^x}{2}\frac{1-\sgn(x-y) (-1)^y}{2} \sgn(x-y),
\end{equation}
\begin{equation}\label{eq4.7b}
A_{i,j}=\sum_{x,y\in\Z} \Phi^t_{i,x} S_{x,y} \Phi_{y,j} = [\Phi^t S \Phi]_{i,j}.
\end{equation}
Then the $n$-point correlation function, for $n\in\{0,\ldots,2N\}$, are given by
\begin{equation}
\rho^{(n)}(x_1,\ldots,x_n)=\Pf\left[K(x_i,x_j)\right]_{i,j=1,\ldots,n}
\end{equation}
where $K$ is a $2\times2$ matrix kernel, $K(x,y)=\left(\begin{array}{cc} K_{1,1}(x,y) &  K_{1,2}(x,y) \\  K_{2,1}(x,y) &  K_{2,2}(x,y)\end{array}\right)$, with
\begin{equation}\label{eq2.26}
\begin{array}{rcl}
K_{1,1}(x,y)&=& -\sum_{i,j=-2N+1}^{0} \Phi^t_{i,x} A^{-1}_{i,j} \Phi^t_{j,y}\\[6pt]
K_{1,2}(x,y)&=& -\sum_{i,j=-2N+1}^{0} \Phi^t_{i,x} A^{-1}_{i,j} [\Phi^t S^t]_{j,y} =-K_{2,1}(y,x)\\[6pt]
K_{2,1}(x,y)&=& -\sum_{i,j=-2N+1}^{0} [\Phi^t S^t]_{i,x} A^{-1}_{i,j} \Phi^t_{j,y}\\[6pt]
K_{2,2}(x,y)&=& S^t_{x,y}-\sum_{i,j=-2N+1}^{0} [\Phi^t S^t]_{i,x} A^{-1}_{i,j} [\Phi^t S^t]_{j,y}.
\end{array}
\end{equation}
\end{proposition}

When $N\to\infty$, (\ref{eqFP}) becomes a Fredholm Pfaffian, $\Pf\left(J - K g\right)=\sqrt{\Det(\Id-J^{-1} K g)}$, where $J=\left(\begin{array}{cc}0&1\\ -1 & 0 \end{array}\right)$, see Section 8 of~\cite{Ra00}. In this case, we consider bounded functions $g$ with support bounded from below, so that the sum in (\ref{eqFP}) is well defined. From the point of view of operators, the determinant has to be though as defined through the modified determinant like in the case of the GOE case, see discussion at the end of Section~\ref{RM}. Finally, note that $A$ is invertible because $\Det(A)$ is the partition function of the line ensemble.

\begin{proof} Since it is often used, we denote the ordered set $I=\{-2N+1,\ldots,0\}$, and instead of writing a matrix $M=[M_{i,j}]_{i,j=-2N+1,\ldots,0}$ we write $M=[M_{i,j}]_{i,j\in I}$. 
Let $w(\{x^{\rm in}\}\to\{x^{\rm fin}_{\bf n}\})$, ${\bf n}\in C_N$ as given in (\ref{eq4.2b}), be the weight of fermions starting from positions $\{x^{\rm in}\}=(x_i^{\rm in})_{i\in I}$, $x_i^{\rm in}=i$, and ending at $\{x^{\rm fin}_{\bf n}\}=(x_j^{\rm fin})_{j\in I}$, $x_j^{\rm fin}=j+2n_j$.
The non-intersection constraint implies~\cite{KM59} that the weight can be expressed via determinants,
\begin{equation}\label{eq7}
w(\{x^{\rm in}\}\to\{x^{\rm fin}_{\bf n}\}) = \Det[\varphi_{i,j}]_{i,j\in I}
\end{equation}
with
\begin{equation}\label{eq8}
\varphi_{i,j}=\bra{\emptyset} a_{j+2n_j} e^{\tilde T \alpha_1} a^*_{i} \ket{\emptyset} = \big[e^{\tilde T \alpha_1}\big]_{j+2n_j,i}=\Phi_{j+2n_j,i}.
\end{equation}
Taking into account the even/odd initial position of the fermions, (\ref{eq7}) can be rewritten as
\begin{equation}\label{eq4.16b}
w(\{x^{\rm in}\}\to\{x^{\rm fin}_{\bf n}\}) = \Det[\Phi_{i}(x_{j}^{\rm fin})]_{i,j\in I}\prod_{j=-N+1}^0 {\bf e}(x^{\rm fin}_{2j}){\bf o}(x^{\rm fin}_{2j-1})
\end{equation}
with
\begin{equation}
{\bf e}(x)=\frac{1+(-1)^x}{2},\quad {\bf o}(x)=\frac{1-(-1)^x}{2}.
\end{equation}

Let us denote by $p(x_{-2N+1},\ldots,x_{0})$ the probability that the set of end points $\{x^{\rm fin}_j,j=-2N+1,\ldots,0\}$ coincide with the set $\{x_{-2N+1},\ldots,x_{0}\}$. We want to show that this probability can be written as a determinant times a Pfaffian. Since the $x_j$'s do not have to be ordered, let $\pi$ be the permutation of $\{-2N+1,\ldots,0\}$ such that $x_{\pi(i)}<x_{\pi(i+1)}$, that is, $x_{\pi(i)}=x^{\rm fin}_{i}$, $i\in I$. Moreover, define the matrix $\Xi=[\Xi_{i,j}]_{i,j\in I}$ by setting $\Xi_{i,j}=\delta_{i,\pi(j)}$. Then
\begin{equation}
[\Phi_{i}(x_{j}^{\rm fin})]_{i,j\in I}=[\Phi_{i}(x_j)]_{i,j\in I} \, \Xi.
\end{equation}
Now let us show that
\begin{equation}\label{eqPfaff}
\prod_{j=-N+1}^0 {\bf e}(x^{\rm fin}_{2j}){\bf o}(x^{\rm fin}_{2j-1})=\Pf[S^t_{x_{i}^{\rm fin},x_{j}^{\rm fin}}]_{i,j\in I}.
\end{equation}
Since $x_{i}^{\rm fin}<x_{i+1}^{\rm fin}$, the components $i,j$ ($i<j$) of the r.h.s.\ matrix are given by 
${\bf o}(x_{i}^{\rm fin}){\bf e}(x_{j}^{\rm fin})$. The Pfaffian of a matrix $M=[M_{i,j}]_{i,j\in I}$ is
\begin{equation}
\Pf(M)=\sum_{\begin{subarray}{c}\sigma\\ \sigma_{2i-1}<\sigma_{2i}\end{subarray}} (-1)^{|\sigma|} \prod_{i=-N+1}^0 M_{\sigma_{2i-1},\sigma_{2i}},
\end{equation}
where the sum is on the permutations $\sigma$ of $\{-2N+1,\ldots,0\}$ with $\sigma_{2i-1}<\sigma_{2i}$. The identity permutation gives already l.h.s.\ of (\ref{eqPfaff}). Thus we have to show that all other terms cancels pairwise. Take a permutation $\sigma$ such that $\sigma(2i-1) < \sigma(2j-1) < \sigma(2i) < \sigma(2j)$ and define the permutation $\sigma'$ by setting $\sigma'(2j)=\sigma(2i)$, $\sigma'(2i)=\sigma'(2j)$, and $\sigma'(k)=\sigma(k)$ otherwise. The term of the Pfaffian coming from $\sigma$ and $\sigma'$ are identical up to a minus sign because $(-1)^{|\sigma|}=-(-1)^{|\sigma'|}$. Moreover, the only permutation for which $\sigma(2i-1) < \sigma(2j-1) < \sigma(2i) < \sigma(2j)$ can not be satisfied for some $i,j$ is the identity. Consequently (\ref{eqPfaff}) holds.

Finally, define the matrix $G=\Xi^t\, [S^t_{x_i,x_j}]_{i,j\in I} \,\Xi$. Replacing the definition of $\Xi$ we obtain $G=[S^t_{x_{i}^{\rm fin},x_{j}^{\rm fin}}]_{i,j\in I}$. Then
\begin{eqnarray}\label{eq4.27T}
& &p(x_{-2N+1},\ldots,x_{0}) = w(\{x^{\rm in}\}\to\{x^{\rm fin}_n\}) \\
&=& \Det[\Phi_{i}(x_{j}^{\rm fin})]_{i,j\in I}\prod_{j=-N+1}^0 {\bf e}(x^{\rm fin}_{2j}){\bf o}(x^{\rm fin}_{2j-1})\nonumber \\
&=& \Det[\Phi_{i}(x_j)]_{i,j\in I} \Det(\Xi) \Pf(\Xi^t\, [S^t_{x_i,x_j}]_{i,j\in I} \,\Xi)\nonumber \\
&=& \Det[\Phi_{i}(x_j)]_{i,j\in I} \Pf[S^t_{x_i,x_j}]_{i,j\in I}\nonumber
\end{eqnarray}
where we used the property of Pfaffians $\Pf(\Xi^t T \Xi)=\Pf(T)\Det(\Xi)$, see e.g.~\cite{Ste90}, and $\Det(\Xi)=(-1)^{|\pi|}$.

The probability (\ref{eq4.27T}) is of the form (\ref{eq3.2}) with
\begin{equation}
\e(x,y)=S^t_{x,y},\quad f_i(x)=\Phi_{x,i}
\end{equation}
from which follows that
\begin{equation}
M_{i,j}=-A_{i,j}, \quad (\e f_i)(x)=-[S \Phi]_{x,i},
\end{equation}
and the kernel is given by
\begin{equation}\label{eq3.14}
K'(x,y)=\left(\begin{array}{cc}-K_{1,1}(x,y) & K_{1,2}(x,y)\\
K_{2,1}(x,y)& -K_{1,2}(x,y)\end{array}\right).
\end{equation}
But $K$ and $K'$ are two equivalent kernels (they give the same correlation functions) since $K'=U^t K U$ with $U=i \left(\begin{array}{cc}1&0\\0& -1 \end{array}\right)$ and $\Pf[U^t K U]=\Det[U]\Pf[K]$. We use $K$ instead of $K'$ uniquely because another derivation of the kernel gave $K$ and we already carried out the analysis.
\end{proof}

\section{Kernel for finite $T$}\label{sect5}
In this section we compute the components of the kernel given in (\ref{eq2.26}). At this stage we take the limit $N\to\infty$. The justification of this limit is in the end of this section. The first step is to find the inverse of the matrix $A$. First we extend $A$ to be defined for all $i,j\in\Z$ by using (\ref{eq4.7b}) to all $i,j$. Let us divide $\ell^2(\Z)=\ell^2(\Z_+^*)\oplus\ell^2(\Z_-)$, where $\Z_+^*=\{1,2,\ldots\}$ and $\Z_-=\{0,-1,\ldots\}$. The inverse of $A$ in (\ref{eq2.26}) is the one in the subspace $\ell^2(\Z_-)$.
Let us denote by $P_-$ the projector on $\Z_-$ and $P_+$ the one on $\Z_+^*$.
\begin{lemma}\label{inverseA}
The inverse of $A$ in subspace $\ell^2(\Z_-)$, which can be expressed as $P_- (P_- A P_- + P_+)^{-1} P_-$, is given by
\begin{equation}
[A^{-1}]_{i,j}=[\alpha_{-1} e^{-\tilde T \alpha_{-1}} P_- e^{-\tilde T \alpha_1}-e^{-\tilde T \alpha_{-1}} P_- e^{-\tilde T \alpha_1} \alpha_1]_{i,j}
\end{equation}
where $[\alpha_1]_{i,j}=\delta_{i,j+1}$ and $\alpha_{-1}\equiv\alpha_1^t$.
\end{lemma}
\begin{proof}
First we rewrite $A$ as a sum of a Toeplitz matrix plus the remainder. Let $\alpha_e$ be the matrix with $[\alpha_e]_{i,j}=\delta_{i,j} \frac{1+(-1)^i}{2}$ and $\alpha_o=\Id-\alpha_e$. Then
\begin{equation}\label{eq4.2}
S=\sum_{k\geq 0} \alpha_1^{2k+1} \alpha_o-\sum_{k\geq 0} \alpha_{-1}^{2k+1} \alpha_e.
\end{equation}
It is then easy to see that, for $V_e(x)$ an even polynomial of arbitrarily high order
\begin{equation}\label{eq4.3}
V_e(\alpha_{\pm 1}) \alpha_e = \alpha_e V_e(\alpha_{\pm 1}),\quad V_e(\alpha_{\pm 1}) \alpha_o = \alpha_o V_e(\alpha_{\pm 1})
\end{equation}
and for $V_o(x)$ an odd polynomial of arbitrarily high order
\begin{equation}\label{eq4.4}
V_o(\alpha_{\pm 1}) \alpha_e = \alpha_o V_o(\alpha_{\pm 1}),\quad V_o(\alpha_{\pm 1}) \alpha_o = \alpha_e V_o(\alpha_{\pm 1}).
\end{equation}
Hence $A$ can be written as
\begin{equation}\label{eq4.5}
A=\exp(\tilde T\alpha_{-1}) \sum_{k\geq 0} (\alpha_1^{2k+1} \alpha_o-\alpha_{-1}^{2k+1}\alpha_e) (\cosh(\tilde T\alpha_1)+\sinh(\tilde T\alpha_1)).
\end{equation}
We pull the last factor in (\ref{eq4.5}) in front of the sum using the commutation relations (\ref{eq4.3}) and (\ref{eq4.4}), and, after some algebraic manipulations, we obtain
\begin{equation}\label{eq5.6}
A= M + R
\end{equation}
where $M=\frac12 \Phi^t (Q-Q^t) \Phi$, $R=\frac12 (Q+Q^t)(\alpha_o-\alpha_e)$, with $Q=\sum_{k\geq 0}\alpha_1^{2k+1}$ and $\Phi=\exp(\tilde T\alpha_1)$. 

Let $B=[\Phi^{-1}]^{t} (\alpha_{-1} P_- - P_- \alpha_1) \Phi^{-1}$. We want to prove that it is the inverse of $A$ in the subspace $\ell^2(\Z_-)$. First notice that $B_{i,j}=0$ if $i\geq 1$ or $j\geq 1$, which implies $[A\cdot B]_{i,j}=[P_- A P_- \cdot B]_{i,j}$ for $i,j \leq 0$. Therefore, for $i,j \leq 0$,
\begin{equation}
[A\cdot B]_{i,j}= [(M+R)\cdot[\Phi^{-1}]^{t} U_0\Phi^{-1}]_{i,j}
\end{equation}
with
\begin{equation}\label{eq4.8}
U_0=\alpha_{-1} P_- - P_- \alpha_1,
\end{equation}
and, expanding $M+R$, we have
\begin{eqnarray}
[A\cdot B]_{i,j} &=& \left[\left(e^{\tilde T\alpha_{-1}} \tfrac{Q-Q^t}{2} e^{\tilde T\alpha_1} + \tfrac{Q+Q^t}{2}(\alpha_o-\alpha_e)\right)\left(e^{-\tilde T\alpha_{-1}} U_0 e^{-\tilde T\alpha_1}\right)\right]_{i,j}\nonumber \\
&=& \left[e^{\tilde T\alpha_1} U_1 e^{-\tilde T\alpha_1}\right]_{i,j}+\left[e^{\tilde T\alpha_{-1}} U_2 e^{-\tilde T\alpha_1}\right]_{i,j}
\end{eqnarray}
where $U_1=\frac12 (Q-Q^t) U_0$ and $U_2=\frac12 (Q+Q^t)(\alpha_o-\alpha_e) U_0$. The components of these matrices are given by
\begin{eqnarray}\label{eq5.10}
\left[U_1\right]_{n,m}&=&\delta_{n,m} \Id_{[n\leq 0]}+\frac12 \delta_{m,0}\sgn(n-1) \frac{1+(-1)^n}{2},\nonumber \\
\left[U_2\right]_{n,m}&=& \frac12 \delta_{m,0} \frac{1+(-1)^n}{2},
\end{eqnarray}
and a simple algebraic computation leads then to $[A\cdot B]_{i,j}=\delta_{i,j}$ for $i,j\leq 0$. Finally, since $A$ and $B$ are antisymmetric, $[B\cdot A]_{i,j}=[A^t\cdot B^t]_{j,i}=[A\cdot B]_{j,i}=\delta_{i,j}$ too. Therefore $B$ is the inverse of $A$ in the subspace $\ell^2(\Z_-)$.
\end{proof}

The second step is to find an explicit expression for the kernel's elements. Using the fact that $[A^{-1}]_{i,j}$ of Lemma~\ref{inverseA} is zero for $i\geq 1$ or $j\geq 1$, we can extend the sum over all $i,j \in \Z$ and obtain
\begin{equation}
\begin{array}{rcl}
K_{1,1}(x,y)&=&-[\Phi A^{-1}\Phi^t]_{x,y},\\[6pt]
K_{1,2}(x,y) &=& -K_{2,1}(y,x),\\[6pt]
K_{2,1}(x,y)&=&-[S \Phi A^{-1} \Phi^t]_{x,y},\\[6pt]
K_{2,2}(x,y)&=& S^t_{x,y}-[S \Phi A^{-1} \Phi^t S^t]_{x,y}.
\end{array}
\end{equation}
Put $\Psi=e^{\tilde T\alpha_1}e^{-\tilde T\alpha_{-1}}$.
We write $S$ as in (\ref{eq4.2}), use the commutation relations
(\ref{eq4.3}) and (\ref{eq4.4}), and after some straightforward algebra obtain
\begin{eqnarray}
K_{1,1} &=& -\Psi U_0 \Psi^t, \nonumber\\
K_{2,1} &=& -\Psi^t (S U_0 - U_1)\Psi^t-\Psi U_1 \Psi^t, \\
K_{2,2} &=& S^t + S K_{1,1}^t, \nonumber
\end{eqnarray}
where $U_1$ is given by (\ref{eq5.10}), and
\begin{eqnarray}
\left[U_0\right]_{n,m}&=& (\delta_{n,m-1}-\delta_{m,n-1})\Id_{[n,m\leq 0]}\nonumber \\
\left[S U_0-U_1\right]_{n,m}&=&\frac12 \frac{1+(-1)^n}{2}\delta_{m,0}.
\end{eqnarray}

Using these relations we obtain the kernel elements, which are summed up in the
following
\begin{lemma}\label{lemma5.2}
\begin{equation}
K(x,y)=G(x,y)+R(x,y),
\end{equation}
with
\begin{eqnarray}
R_{1,1}(x,y)&=&0,\nonumber\\
R_{1,2}(x,y)&=&-\frac{(-1)^{y}}{2}J_{x+1}(2\tilde T),\nonumber\\
R_{2,1}(x,y)&=&\frac{(-1)^{x}}{2}J_{y+1}(2\tilde T),\nonumber\\
R_{2,2}(x,y)&=&-S(x,y)+\frac14 \sgn(x-y)\nonumber \\
& &-\frac{(-1)^x}{2}\sum_{m\geq 1}J_{y+2m}(2\tilde T)+\frac{(-1)^{y}}{2}\sum_{n\geq 1}J_{2n+x}(2\tilde T),
\end{eqnarray}
and
\begin{equation}
G_{1,1}(x,y)=-\sum_{n\geq 1} J_{x+n+1}(2\tilde T) J_{y+n}(2\tilde T)+\sum_{n\geq 1} J_{y+n+1}(2\tilde T) J_{x+n}(2\tilde T),
\end{equation}
\begin{equation}
G_{1,2}(x,y)=\sum_{n\geq 1} J_{x+n}(2\tilde T) J_{y+n}(2\tilde T) - J_{x+1}(2\tilde T)\bigg( \sum_{m\geq 1} J_{y+2m-1}(2\tilde T)-\frac12\bigg),
\end{equation}
\begin{equation}
G_{2,1}(x,y)=-\sum_{n\geq 1} J_{x+n}(2\tilde T) J_{y+n}(2\tilde T) + J_{y+1}(2\tilde T)\bigg( \sum_{m\geq 1} J_{x+2m-1}(2\tilde T)-\frac12\bigg),
\end{equation}
\begin{eqnarray}
G_{2,2}(x,y)&=&\sum_{m\geq 1}\sum_{n\geq m}J_{x+2m}(2\tilde T) J_{y+2n+1}(2\tilde T)-\sum_{n\geq 1}\sum_{m\geq n} J_{x+2m+1}(2\tilde T) J_{y+2n}(2\tilde T)\nonumber \\
& &-\frac12 \sum_{m\geq 1} J_{x+2m}(2\tilde T)+\frac12 \sum_{n\geq 1} J_{y+2n}(2\tilde T)-\frac14 \sgn(x-y),
\end{eqnarray}
where $J_m(t)$ denotes the m$^{\rm th}$ order Bessel function.
\end{lemma}

Remark: this result could also be deduced starting from Section 5 of~\cite{Ra00}. 
Now we justify the $N\to\infty$ limit. Let us first explain the idea. Denote the sets $I=\{-2N+1,\ldots,0\}$ and $L=\{-N+1,\ldots,0\}$. We consider the kernel's elements for $x,y\geq 0$. For $(i,j)\in I^2\setminus L^2$, the inverse of $A$ for finite $N$ differs from the inverse for $N=\infty$ only by $\Or(e^{-\mu N})$ with $\mu=\mu(\tilde T)>0$. On the other hand, the contribution to $K_{.,.}(x,y)$ coming from $(i,j)\in (I\setminus L)^2$ are exponentially small in $N$. Therefore, replacing the inverse of $A$ for finite $N$ with the inverse obtained in Lemma~\ref{inverseA} we introduce only an error exponentially small in $N$. The dependence of the kernel's elements on $N$ is only via the extension of the sums in (\ref{eq2.26}), which limit the one we derived in Lemma~\ref{lemma5.2}.

In what follows we denote by $A_N$ the $2N\times 2N$ matrix (\ref{eq4.7b}) and by $A$ the $N=\infty$ one.
\begin{lemma}
If we replace $[A_N^{-1}]_{i,j}$ by $A^{-1}_{i,j}$ in the kernel's elements (\ref{eq2.26}), then for $N$ large enough, the error made is $\Or(e^{-\mu N})$ for some constant $\mu=\mu(\tilde T)>0$. The error is uniform for $x,y\geq 0$.
\end{lemma}
\begin{proof}
Here we use some results of Appendix~\ref{AppAAA}. First, we define the matrix $B$ by setting, $B_{i,j}=A^{-1}_{i,j}$ for $(i,j)\in I\times L$, and $B_{i,j}=-A^{-1}_{-2N+1-i,-2N+1-j}$ for $(i,j)\in I\times (I\setminus L)$. Since $[A_N]_{i,j}=-[A_N]_{-2N+1-i,-2N+1-j}$, by (\ref{eqAppA7}) follows that
\begin{equation}
A_N B = \Id -C
\end{equation}
for some matrix $C$ with $\|C\|=\max_{i,j} |C_{i,j}| \leq \Or(e^{-\mu_2 N})$. Therefore, for $N$ large enough,
\begin{equation}
A_N^{-1} = B (\Id +D),\quad D=\sum_{k\geq 1} C^k
\end{equation}
with $\|D\| \leq \Or(e^{-\mu_2 N})$ too. Thus, replacing $A_N^{-1}$ with $B$ we introduce an error in the kernel's elements of $\Or(N^2 e^{-\mu_2 N})$.

If we replace $B_{i,j}$ with $A^{-1}_{i,j}$ also in $(i,j)\in L\times (I\setminus L)$ we introduce an error of $\Or(N^2 e^{-\mu_3 N})$, with $\mu_3=\min\{\mu_1,\mu_2/2\}$. This is achieved using (\ref{eqAppA6}) for $i<j+N/2$, and (\ref{eqAppA4}) otherwise.

The final step is to show, using only the antisymmetry of $A_N^{-1}$ that the contribution of $K_{.,.}$ coming from $(i,j)\in (I\setminus L)^2$ are also exponentially small in $N$. For $(i,j) \in (I\setminus L)^2$, it is easy to see that, uniformly in $x,y\geq 0$,
\begin{eqnarray}
\Phi_{x,i}&=&\Or(e^{-\mu_1 N}) \\
(S \Phi)_{x,i} &=& \Or(e^{-\mu_1 N}),\textrm{ for odd }x,\nonumber \\
(S \Phi)_{x,i} &=& \sinh(\tilde T)+\Or(e^{-\mu_1 N}),\textrm{ for even }x\textrm{ and even }i,\nonumber\\
(S \Phi)_{x,i} &=& \cosh(\tilde T)+\Or(e^{-\mu_1 N}),\textrm{ for even }x\textrm{ and odd }i.\nonumber
\end{eqnarray}
Therefore, the contributions for $K_{1,1}$, $K_{1,2}$, and $K_{2,1}$ are $\Or(N^2 e^{-\mu_1 N})$ because they contain at least a factor $\Or(e^{-\mu_1 N})$ coming from $\Phi_{x,i}$ or $\Phi^t_{j,y}$. For $K_{2,2}$ there are terms without $\Or(e^{-\mu_1 N})$, and containing only $\sinh(\tilde T)$ and/or $\cosh(\tilde T)$. These terms cancel exactly because $A_N^{-1}$ is antisymmetric. Consequently, we can simply replace $B_{i,j}$ with $A^{-1}_{i,j}$ also in $(i,j)\in (I\setminus L)^2$ up to an error $\Or(N^2 e^{-\mu_1 N})$.
\end{proof}

\section{Edge scaling and asymptotics of the kernel}\label{sect6}
In this section we define the edge scaling of the kernel, provide some bounds on them which will be used in the proofs of Section~\ref{sect7}, and compute their $T\to\infty$ limit.

The edge scaling of the kernel is defined by
\begin{eqnarray}\label{eq5.19}
G_{T;1,1}^{\mathrm{edge}}(\xi_1,\xi_2) &=& \tilde T^{2/3} G_{1,1}([2\tilde T+\xi_1 \tilde T^{1/3}],[2\tilde T+\xi_2 \tilde T^{1/3}]) \nonumber \\
G_{T;k}^{\mathrm{edge}}(\xi_1,\xi_2) &=& \tilde T^{1/3} G_{k}([2\tilde T+\xi_1 \tilde T^{1/3}],[2\tilde T+\xi_2 \tilde T^{1/3}]),\quad k=(1,2),(2,1) \nonumber\\
G_{T;2,2}^{\mathrm{edge}}(\xi_1,\xi_2) &=& G_{2,2}([2\tilde T+\xi_1 \tilde T^{1/3}],[2\tilde T+\xi_2 \tilde T^{1/3}]),
\end{eqnarray}
and similarly for $R_{T;k}^{\mathrm{edge}}(\xi_1,\xi_2)$.

Next we compute some bounds on the kernel's elements such that, when possible, they are rapidly decreasing for $\xi_1,\xi_2 \gg 1$.

\begin{lemma}\label{LemmaBounds} Write
\begin{equation}
\Omega_0(x)=\left\{\begin{array}{ll}1,&x\leq 0\\
\exp(-x/2),&x\geq 0\end{array}\right. ,\quad \Omega_1(x)=\left\{\begin{array}{ll}1+|x|,&x\leq 0\\
\exp(-x/2),&x\geq 0\end{array}\right. ,
\end{equation}
\begin{equation}
\Omega_2(x)=\left\{\begin{array}{ll}(1+|x|)^2,&x\leq 0\\
\exp(-x/2),&x\geq 0\end{array}\right. .
\end{equation}
Then there is a positive constant $C$ such that for large $\tilde T$
\begin{eqnarray}
|R_{T;1,2}^{\mathrm{edge}}(\xi_1,\xi_2)| &\leq & C \Omega_0(\xi_1),\nonumber \\
|R_{T;2,1}^{\mathrm{edge}}(\xi_1,\xi_2)| &\leq & C \Omega_0(\xi_2),\\
|R_{T;2,2}^{\mathrm{edge}}(\xi_1,\xi_2)| &\leq & C (\Omega_1(\xi_1)+\Omega_1(\xi_2)),\nonumber
\end{eqnarray}
and
\begin{eqnarray}
|G_{T;1,1}^{\mathrm{edge}}(\xi_1,\xi_2)| &\leq & C \Omega_2(\xi_1)\Omega_2(\xi_2),\nonumber \\
|G_{T;1,2}^{\mathrm{edge}}(\xi_1,\xi_2)| &\leq & C \Omega_1(\xi_1)(1+\Omega_2(\xi_2)),\\
|G_{T;2,1}^{\mathrm{edge}}(\xi_1,\xi_2)| &\leq & C \Omega_1(\xi_2)(1+\Omega_2(\xi_1)), \nonumber \\
|G_{T;2,2}^{\mathrm{edge}}(\xi_1,\xi_2)| &\leq & C (1+\Omega_1(\xi_1)+\Omega_1(\xi_2) +\Omega_1(\xi_1)\Omega_1(\xi_2)). \nonumber
\end{eqnarray}
\end{lemma}
\begin{proof}
We use Lemma~\ref{LemmaBessel1} and Lemma~\ref{LemmaBessel2} to obtain the above estimate.

\noindent 1) The bounds on $|R_{T;1,2}^{\mathrm{edge}}(\xi_1,\xi_2)|$ and $|R_{T;2,1}^{\mathrm{edge}}(\xi_1,\xi_2)|$ are implied by Lemma~\ref{LemmaBessel1}.

\noindent 2) Bound on $|R_{T;2,2}^{\mathrm{edge}}(\xi_1,\xi_2)|$.
\begin{equation}
|R_{T;2,2}^{\mathrm{edge}}(\xi_1,\xi_2)| \leq \frac54 + \frac12 \sum_{M\in \N/\tilde T^{1/3}}|J_{[2\tilde T+(2M+\xi_2)\tilde T^{1/3}]}(2\tilde T)|+ (\xi_1\leftrightarrow\xi_2)
\end{equation}
and
\begin{equation}\label{eq6.7}
\sum_{M\in \N/\tilde T^{1/3}}|J_{[2\tilde T+(2M+\xi_2)\tilde T^{1/3}]}(2\tilde T)| \leq \sum_{M\in \N/\tilde T^{1/3}}|J_{[2\tilde T+(M+\xi_2)\tilde T^{1/3}]}(2\tilde T)|.
\end{equation}
For $\xi_2\leq 0$,
\begin{equation}
(\ref{eq6.7})\leq \sum_{M\in \xi_2+\N/\tilde T^{1/3}\cup[\xi_2,0]} |J_{[2\tilde T+M\tilde T^{1/3}]}(2\tilde T)| +\sum_{M\in \N/\tilde T^{1/3}} |J_{[2\tilde T+M\tilde T^{1/3}]}(2\tilde T)|.
\end{equation}
By (\ref{eq5.2b}) the first term is bounded by a constant times $(1+|\xi_2|)$ and by (\ref{eq5.2c}) the second term by a constant. For $\xi_2\geq 0$,
\begin{equation}
(\ref{eq6.7}) \leq \sum_{M\in \xi_2+\N/\tilde T^{1/3}\cup[\xi_2,\infty)} |J_{[2\tilde T+M\tilde T^{1/3}]}(2\tilde T)|
\end{equation}
which, by (\ref{eq5.2c}), is bounded by a constant times $\exp(-\xi_2/2)$. Therefore
\begin{equation}\label{eq5.10b}
\sum_{M\in \N/\tilde T^{1/3}}|J_{[2\tilde T+(M+\xi_2)\tilde T^{1/3}]}(2\tilde T)| \leq C \, \Omega_1(\xi_2).
\end{equation}
for a constant $C$, from which follows the desired bound.

\noindent 3) Bound on $|G_{T;1,1}^{\mathrm{edge}}(\xi_1,\xi_2)|$. Let us define $\tilde J_n(t)=J_{n+1}(t)-J_n(t)$. Then
\begin{eqnarray}\label{eqG1}
G_{T;1,1}^{\mathrm{edge}}(\xi_1,\xi_2) 
&=& \tilde T^{2/3} \sum_{M\in \N/\tilde T^{1/3}} J_{[2\tilde T+(\xi_1+M) \tilde T^{1/3}]}(2\tilde T) \tilde J_{[2\tilde T+(\xi_2+M) \tilde T^{1/3}]}(2\tilde T)\nonumber\\ &-& (\xi_1\leftrightarrow \xi_2).
\end{eqnarray}
For large $\tilde T$, the sums are very close integrals and this time we use both Lemma~\ref{LemmaBessel1} and Lemma~\ref{LemmaBessel2}, obtaining
\begin{eqnarray}\label{eq5.12b}
|G_{T;1,1}^{\mathrm{edge}}(\xi_1,\xi_2) |
&\leq& C \int_0^\infty \dx M \Omega_0(M+\xi_1)\Omega_1(M+\xi_2) \nonumber \\
&\leq& C \int_0^\infty \dx M \Omega_1(M+\xi_1)\Omega_1(M+\xi_2)
\end{eqnarray}
for a constant $C>0$.
It is then easy to see that r.h.s.\ of (\ref{eq5.12b}) is bounded as follows: for $\xi_1\leq\xi_2\leq 0$ by $C (1+|\xi_1|)^2$, for $\xi_1 \leq 0\leq \xi_2$ by $C (1+|\xi_1|)^2 \exp(-\xi_2/2)$, and for $0\leq \xi_1\leq \xi_2$ by $C \exp(-\xi_1/2)\exp(-\xi_2/2)$, for some other constant $C>0$. Therefore $|G_{T;1,1}^{\mathrm{edge}}(\xi_1,\xi_2) |\leq C \Omega_2(\xi_1)\Omega_2(\xi_2).$

\noindent 4) Bound on $|G_{T;1,2}^{\mathrm{edge}}(\xi_1,\xi_2)|$. 
\begin{eqnarray*}
& &G_{T;1,2}^{\mathrm{edge}}(\xi_1,\xi_2) 
= \sum_{M\in \N/\tilde T^{1/3}} J_{[2\tilde T+(\xi_1+M) \tilde T^{1/3}]}(2\tilde T) \big(T^{1/3} J_{[2\tilde T+(\xi_2+M) \tilde T^{1/3}]}(2\tilde T)\big)\\ & &- 
T^{1/3} J_{[2\tilde T+(\xi_1+M) \tilde T^{1/3}+1]}(2\tilde T)
\Big( \sum_{M\in \N/\tilde T^{1/3}} J_{[2\tilde T+(\xi_1+2M) \tilde T^{1/3}-1]}(2\tilde T) -\frac12 \Big).
\end{eqnarray*}
In the first sum, the term with $\xi_2$ is bounded by a constant and remaining sum was already estimated in (\ref{eq5.10b}). The second term is bounded by a constant times $\Omega_0(\xi_1)\Omega_2(\xi_2)$. Using $\Omega_0(\xi_1)\leq\Omega_1(\xi_1)$ we conclude that
$|G_{T;1,2}^{\mathrm{edge}}(\xi_1,\xi_2)|\leq C \Omega_1(\xi_1)(1+\Omega_2(\xi_2)).$

\noindent 5) Bound on $|G_{T;2,1}^{\mathrm{edge}}(\xi_1,\xi_2)|$. The bound is the same as for $|G_{T;1,2}^{\mathrm{edge}}(\xi_1,\xi_2)|$.

\noindent 6) Bound on $|G_{T;2,2}^{\mathrm{edge}}(\xi_1,\xi_2)|$. 
The terms with the double sums are estimated applying twice (\ref{eq5.10b}) and are then bounded by $\Omega_1(\xi_1)\Omega_1(\xi_2)$. The two terms with only one sum are bounded by $\Omega_1(\xi_1)$ and $\Omega_1(\xi_2)$ respectively, and the signum function by $1/4$. Therefore, for some constant $C>0$,
$|G_{T;2,2}^{\mathrm{edge}}(\xi_1,\xi_2)|\leq C (1+\Omega_1(\xi_1)+\Omega_1(\xi_2)+\Omega_1(\xi_1)\Omega_1(\xi_2)).$
\end{proof}

Finally we compute the pointwise limits of the $G$'s since they remains in the weak convergence.
\begin{lemma} For any fixed $\xi_1,\xi_2$,
\begin{equation}
\lim_{\tilde T\to\infty} G_{T;k}^{\mathrm{edge}}(\xi_1,\xi_2) = G_k^{\rm GOE}(\xi_1,\xi_2),
\end{equation}
where the $G_k^{\rm GOE}$'s are the ones in (\ref{eqGOEkernel}).
\end{lemma}
\begin{proof}
Let us consider $\xi_1,\xi_2$ fixed. In the proof of Lemmas~\ref{LemmaBounds}, we have already obtained uniform bounds in $T$ for $G_{T;k}^{\mathrm{edge}}(\xi_1,\xi_2)$, so that dominated convergence applies. To obtain the limits we use (\ref{Bessel2Airy}), i.e.,
\begin{equation}
\lim_{T\to\infty}T^{1/3}J_{[2T+\xi T^{1/3}]}(2T)=\Ai(\xi),
\end{equation}
and
\begin{equation}
\lim_{T\to\infty}T^{2/3}(J_{[2T+\xi T^{1/3}+1]}(2T)-J_{[2T+\xi T^{1/3}]}(2T))=\Aip(\xi).
\end{equation}
The limit of $G_{T;1,1}^{\mathrm{edge}}(\xi_1,\xi_2)$ follows from (\ref{eqG1}).\\
The limit of $G_{T;1,2}^{\mathrm{edge}}(\xi_1,\xi_2)$ leads to
\begin{equation}
\int_0^{\infty} \dx\lambda \Ai(\xi_1+\lambda)\Ai(\xi_2+\lambda)-\frac12\Ai(\xi_1) \bigg(\int_0^{\infty}\dx\lambda \Ai(\xi_2+\lambda)-1\bigg)
\end{equation}
which equals $G_{1,2}^{\rm GOE}$ since $\int_0^{\infty}\dx\lambda \Ai(\xi_2+\lambda)-1=-\int_0^{\infty}\dx\lambda \Ai(\xi_2-\lambda)$.\\
The limit of $G_{T;2,1}^{\mathrm{edge}}(\xi_1,\xi_2)$ is obtained identically.\\
Finally, the limit of $G_{T;2,2}^{\mathrm{edge}}(\xi_1,\xi_2)$ is given by
\begin{eqnarray}
& &\frac14 \int_0^\infty\dx\lambda\int_\lambda^\infty\dx\mu \Ai(\xi_1+\lambda)\Ai(\xi_2+\mu) - (\xi_1\leftrightarrow\xi_2) \\
&-&\frac14 \int_0^\infty\dx\lambda \Ai(\xi_1+\lambda)+\frac14 \int_0^\infty\dx\mu \Ai(\xi_2+\mu)-\frac14\sgn(\xi_1-\xi_2),\nonumber
\end{eqnarray}
which can be written in a more compact form.
Since $\int_\R\dx\lambda\Ai(\lambda)=1$,
\begin{equation}\label{eq5.38}
\int_0^\infty\dx\lambda \Ai(\xi_1+\lambda) = \int_0^\infty\dx\lambda\int_{-\infty}^\infty\dx\mu \Ai(\xi_1+\lambda) \Ai(\xi_2+\mu),
\end{equation}
and the signum can be expressed as an integral of $\Ai(\xi_1+\lambda) \Ai(\xi_2+\mu)$
\begin{equation}\label{eq5.39}
-\sgn(\xi_1-\xi_2)=\int_\R\dx\lambda\int_\R\dx\mu \Ai(\xi_1+\lambda) \Ai(\xi_2+\mu) \sgn(\lambda-\mu).
\end{equation}
In fact
\begin{equation}
\textrm{r.h.s.\ of }(\ref{eq5.39}) = \int_\R\dx\lambda\int_\R\dx\mu \Ai(\lambda) \Ai(\mu) \sgn(\lambda-\mu+\zeta) = b(\zeta)
\end{equation}
with $\zeta=\xi_2-\xi_1$. For $\zeta=0$ it is zero by symmetry. Then consider $\zeta>0$, the case $\zeta<0$ follows by symmetry. By completeness of the Airy functions,
\begin{equation}
\frac{\dx b(\zeta)}{\dx\zeta}= \int_\R\dx\mu\Ai(\mu)\Ai(\mu-\zeta) = \delta(\zeta).
\end{equation}
Then using (\ref{eq5.38}) and (\ref{eq5.39}) we obtain the result.
\end{proof}

Remark that the GOE kernel in~\cite{SI03} differs slightly from the one written here, but they are equivalent in the sense that they give the same correlation functions.

For the residual terms the limit does not exist, but exists in the even/odd positions. In particular
\begin{equation}
\lim_{T\to\infty}\sum_{m\geq 1}J_{[2\tilde T+\xi \tilde T^{1/3}+2m]}(2\tilde T) = \frac12 \int_0^\infty\dx\lambda \Ai(\xi+\lambda).
\end{equation}

\section{Proof of Theorem~\ref{thmMain}}\label{sect7}
In this section we first prove the weak convergence of the edge rescaled point process of $\eta_T^{\mathrm{sym}}$ to $\eta^{\rm GOE}$ in the $T\to\infty$ limit. Secondly, using the equivalence of the point process $\zeta_T^{\mathrm{sym}}$ and $\zeta_T^{\mathrm{flat}}$, we prove Theorem~\ref{thmMain}.

\begin{theorem}\label{thm0} Let us define the rescaled point process
\begin{equation}
\eta_T^{\rm sym}(f) = \sum_{x \in \Z}f((x-2\tilde T)/\tilde T^{1/3}) \zeta_T^{\rm sym}(x)
\end{equation}
with $\tilde T=\sqrt{2\varrho} T = 2T$ and $f$ a smooth test function of compact support. In the limit $T\to\infty$ it converges weakly to the GOE point process, i.e., for all $m \in \N$, and $f_1,\ldots,f_m$ smooth test functions of compact support,
\begin{equation}
\lim_{T\to\infty} \E_T\bigg(\prod_{k=1}^m\eta_T^{\rm sym}(f_k)\bigg) = \E\bigg(\prod_{k=1}^m\eta^{\rm GOE}(f_k)\bigg)
\end{equation}
where the GOE kernel is given in~\ref{eqGOEkernel}.
\end{theorem}
\begin{proof}
Let $f_1,\ldots,f_m$ be smooth test functions of compact support and $\hat f_i(x)=f_i((x-2\tilde T)/\tilde T^{1/3})$, then
\begin{eqnarray}
& &\E_T\bigg(\prod_{k=1}^m\eta_T^{\rm sym}(f_k)\bigg)=\sum_{x_1,\ldots,x_m\in\Z} \hat f_1(x_1)\ldots \hat f_m(x_m) \Pf[K(x_i,x_j)]_{i,j=1,\ldots,m}\nonumber\\
&=&\sum_{x_1,\ldots,x_m\in\Z} \hat f_1(x_1)\ldots \hat f_m(x_m) \Pf[(X K X^t)(x_i,x_j)]_{i,j=1,\ldots,m}/\Det[X]^m\nonumber\\
&=&\frac{1}{\tilde T^{m/3}}\sum_{x_1,\ldots,x_m\in\Z} \hat f_1(x_1)\ldots \hat f_m(x_m) \Pf[L(x_i,x_j)]_{i,j=1,\ldots,m}
\end{eqnarray}
where $X=\left(\begin{array}{cc}\tilde T^{1/3}&0\\0& 1\end{array}\right)$ and $L(x,y)=(X K X^t)(x,y)$, i.e., 
$L_{1,1}(x,y)=\tilde T^{2/3} K_{1,1}(x,y)$, $L_k(x,y)=\tilde T^{1/3} K_k(x,y)$, for $k=(1,2),(2,1)$, and $L_{2,2}(x,y)= K_{2,2}(x,y)$.
Moreover, we define the edge scaling for the kernel elements as
\begin{equation}
L^{\mathrm{edge}}_{T;k}(\xi_1,\xi_2)=L_k([2\tilde T+\xi_1 \tilde T^{1/3}],[2\tilde T+\xi_2 \tilde T^{1/3}]).
\end{equation}
In what follows we denote by $\xi_i=(x_i-2\tilde T)/\tilde T^{1/3}$. To simplify the notations we consider $\tilde T\in \N$, but the same proof can be carried out without this condition, replacing for example $\Z/\tilde T^{1/3}$ by $(\Z-2\tilde T)/\tilde T^{1/3}$ in (\ref{eq6.4}).
Then
\begin{equation}\label{eq6.4}
\E_T\bigg(\prod_{k=1}^m\eta_T^{\rm sym}(f_k)\bigg)=\frac{1}{\tilde T^{m/3}}\hspace{-0.5cm}\sum_{\xi_1,\ldots,\xi_m\in \Z/\tilde T^{1/3}} \hspace{-0.5cm}f_1(\xi_1)\cdots f_m(\xi_m) \Pf[L^{\mathrm{edge}}_T(\xi_i,\xi_j)]_{i,j=1,\ldots,m}.
\end{equation}
Let us denote $\xi_i^I=[\xi_i \tilde T^{1/3}]/\tilde T^{1/3}$ the ``integer'' discretization of $\xi_i$. Then
\begin{equation}\label{eq6.6}
\E_T\bigg(\prod_{k=1}^m\eta_T^{\rm sym}(f_k)\bigg)=\int_{\R^m}\dx\xi_1\cdots\dx\xi_m \hspace{-0.2cm}f_1(\xi_1^I)\cdots f_m(\xi_m^I) \Pf[L^{\mathrm{edge}}_T(\xi_i^I,\xi_j^I)]_{i,j=1,\ldots,m}.
\end{equation}
Using the definition in (\ref{eq5.19}) we have
\begin{equation}
L^{\mathrm{edge}}_{T;k}(\xi_1,\xi_2)=G^{\mathrm{edge}}_{T;k}(\xi_1,\xi_2)+R^{\mathrm{edge}}_{T;k}(\xi_1,\xi_2),
\end{equation}
therefore (\ref{eq6.6}) consists in one term with only $G^{\mathrm{edge}}_{T;k}$ plus other terms which contain at least one $R^{\mathrm{edge}}_{T;k}$. 

First consider the contribution where only $G^{\mathrm{edge}}_{T;k}$ occur.
Let $M_f>0$ be the smallest number such that $f_j(x)=0$ if $|x|\geq M_f$, for all $j=1,\ldots,m$. We bound the product of the $f_i$'s by
\begin{equation}\label{eq7.8}
|f_1(\xi_1^I)\cdots f_m(\xi_m^I)| \leq \prod_{j=1}^m\|f_j\|_{\infty} \Id_{[-M_f,M_f]}(\xi_j)
\end{equation}
and, in the same way as in Lemma~\ref{lemma6.3} but with $K_{T;k}^{\textrm{edge}}$ replaced by $G_{T;k}^{\textrm{edge}}$, we conclude that this is uniformly integrable in $T$. We then apply dominated convergence and take the limit inside the integral obtaining
\begin{eqnarray}
& &\lim_{T\to\infty}\int_{\R^m}\dx\xi_1\cdots\dx\xi_m f_1(\xi_1^I)\cdots f_m(\xi_m^I) \Pf[G^{\mathrm{edge}}_T(\xi_i^I,\xi_j^I)]_{i,j=1,\ldots,m}\nonumber \\
& = & \int_{\R^m}\dx\xi_1\cdots\dx\xi_m f_1(\xi_1)\cdots f_m(\xi_m) \Pf[G^{GOE}(\xi_i,\xi_j)]_{i,j=1,\ldots,m}.
\end{eqnarray}

Next we have to show that whenever some $R^{\mathrm{edge}}_{T;k}$ are present their contribution vanish in the limit $T\to\infty$. In (\ref{eq6.6}) we have to compute the Pfaffian of $E_T$ defined by
\begin{equation}
E_T(n,l) = \left\{\begin{array}{ll} L^{\mathrm{edge}}_{T;1,1}((n+1)/2,(l+1)/2),& n\textrm{ odd}, l\textrm{ odd},\\ L^{\mathrm{edge}}_{T;1,2}((n+1)/2,l/2), & n\textrm{ odd}, l\textrm{ even},\\ L^{\mathrm{edge}}_{T;2,1}(n/2,(l+1)/2), & n\textrm{ even}, l\textrm{ odd},\\ L^{\mathrm{edge}}_{T;2,2}(n/2,l/2), & n\textrm{ even}, l\textrm{ even},\end{array}\right.
\end{equation}
for $1\leq n < l \leq 2m$, with $L^{\mathrm{edge}}_{T;k}(a,b)\equiv L^{\mathrm{edge}}_{T;k}(\xi_a,\xi_b)$. The Pfaffian of $E_T$ is given by
\begin{equation}\label{eq4.36}
\Pf(E_T)=\sum_{\begin{subarray}{c}\sigma\in S_{2m}\\ \sigma_{2i-1}<\sigma_{2i}\end{subarray}} (-1)^{|\sigma|} E_T(\sigma_1,\sigma_2)\cdots E_T(\sigma_{2m-1},\sigma_{2m}).
\end{equation}
Now we have to check that the product of residual terms does not contain twice the term $(-1)^x$ for the same $x$. This is implied by Lemma~\ref{lemmatwice}.

Let us decompose the sum in (\ref{eq6.4}) into $2^m$ sums, depending on whether $\xi_i \tilde T^{1/3}$ is even or odd. Denote $\xi_i^e=[\xi_i \tilde T^{1/3}/2] 2/\tilde T^{1/3}$ and $\xi_i^o=([\xi_i \tilde T^{1/3}/2] 2+1)/\tilde T^{1/3}$ the ``even'' and ``odd'' discretizations of $\xi_i$. Then
\begin{equation}
(\ref{eq6.4})=\frac{1}{2^m}\sum_{\begin{subarray}{c}s_i=\{o,e\},\\i=1,\ldots,m\end{subarray}} \int_{\R^m}\dx\xi_1\cdots\dx\xi_m \hspace{-0.3cm}f_1(\xi_1^{s_1})\cdots f_m(\xi_m^{s_m})\Pf[L^{\mathrm{edge}}_T(\xi_i^{s_i},\xi_j^{s_j})]_{i,j=1,\ldots,m}.
\end{equation}
With this subdivision, each term in the Pfaffian converges pointwise to a well defined limit. Moreover all the $2^m$ integrals, including $G^{\mathrm{edge}}_{T;k}$'s and/or $R^{\mathrm{edge}}_{T;k}$'s, are uniformly bounded in $T$. By dominated convergence we can take the limit inside the integrals.

Each time that there is a $R^{\mathrm{edge}}_{T;1,2}(\xi_i,\xi_j)$, or $R^{\mathrm{edge}}_{T;2,1}(\xi_j,\xi_i)$, the integral with $s_i=o$ and the one with $s_i=e$ only differs by sign, therefore they cancel each other.
Each time that appears $R^{\mathrm{edge}}_{T;2,2}(\xi_i,\xi_j)$, the part including coming from the $(-1)^{x_i}$ and the one with $(-1)^{x_j}$ simplifies in the same way. Finally we consider the second part, the one including the $S$ and signum function. The sum of $s_i=\{o,e\}$ and $s_j=\{o,e\}$ of the terms with $-S(\xi_i \tilde T^{-1/3},\xi_j \tilde T^{-1/3})$ equals minus the ones with $\tfrac14 \sgn((\xi_i-\xi_j) \tilde T^{-1/3})$. Consequently all the terms including at least one time $R^{\mathrm{edge}}_{T;i}$ have a contribution which vanishes in the $T\to\infty$ limit.
\end{proof}

\begin{lemma}\label{lemmatwice}
The following products do not appear in (\ref{eq4.36}):
\begin{equation}
\begin{array}{ll}
(a) L^{\mathrm{edge}}_{T;2,2}(x_i,x_j) L^{\mathrm{edge}}_{T;1,2}(x_k,x_i), &
(b) L^{\mathrm{edge}}_{T;2,2}(x_i,x_j) L^{\mathrm{edge}}_{T;1,2}(x_k,x_j),\\
(c) L^{\mathrm{edge}}_{T;2,2}(x_i,x_j) L^{\mathrm{edge}}_{T;2,1}(x_i,x_k), &
(d) L^{\mathrm{edge}}_{T;2,2}(x_i,x_j) L^{\mathrm{edge}}_{T;2,1}(x_j,x_k)\\
(e) L^{\mathrm{edge}}_{T;1,2}(x_i,x_j) L^{\mathrm{edge}}_{T;2,1}(x_j,x_k).
\end{array}
\end{equation}
\end{lemma}
\begin{proof} We prove it by reduction ab absurdum. We assume that the product appear and we obtain a contradiction. (a) appears if there exist some $a<b$ and $c<d$ with $a,b,d$ even and $c$ odd, all different, such that $i=a/2$, $j=b/2$, $k=(c+1)/2$, $i=d/2$. But this is not possible since $d\neq a$. (b) appears if there exist some $a<b$ and $c<d$ with $a,b,d$ even and $c$ odd, all different, such that $i=a/2$, $j=b/2$, $k=(c+1)/2$, $j=d/2$. But this is not possible since $d\neq b$. (c) appears if there exist some $a<b$ and $c<d$ with $a,b,c$ even and $d$ odd, all different, such that $i=a/2$, $j=b/2$, $i=c/2$, $k=(d+1)/2$. But this is not possible since $c\neq a$. (d) appears if there exist some $a<b$ and $c<d$ with $a,b,c$ even and $d$ odd, all different, such that $i=a/2$, $j=b/2$, $j=c/2$, $k=(d+1)/2$. But this is not possible since $c\neq b$. (e) appears if there exist some $a<b$ and $c<d$ with $b,c$ even and $a,d$ odd, all different, such that $i=(a+1)/2$, $j=b/2$, $j=c/2$, $k=(d+1)/2$. But this is not possible since $c\neq b$.
\end{proof}

\begin{lemma}\label{lemma6.3}
There exists a constant $C>0$ such that
\begin{equation}
\E_T\Big(|\eta_T^{\rm sym}(\Id_{[-M,\infty)})|^m\Big) \leq C^m e^{M m/2}(m)^{m/2}
\end{equation}
uniformly in $T$.
\end{lemma}
\begin{proof}
The $m$-point correlation function $\rho^{(m)}(\xi_1,\ldots,\xi_m)$ is a sum of product of $K_{T;k}^{\textrm{edge}}$'s which contains twice every $\xi_i$'s, $i=1,\ldots,m$, and only in $K_{T;k}^{\textrm{edge}}$ the two argument can be the same. From Lemma~\ref{LemmaBounds}, for any $\xi_1,\xi_2\in\R$,
\begin{eqnarray}
|K_{T;1,1}^{\textrm{edge}}(\xi_1,\xi_2)| &\leq& C \exp(-\xi_1/2)\exp(-\xi_2/2) \\
|K_{T;1,2}^{\textrm{edge}}(\xi_1,\xi_2)| &\leq& C \exp(-\xi_1/2)\nonumber \\
|K_{T;2,1}^{\textrm{edge}}(\xi_1,\xi_2)| &\leq& C \exp(-\xi_2/2)\nonumber \\
|K_{T;2,2}^{\textrm{edge}}(\xi_1,\xi_2)| &\leq& C.\nonumber 
\end{eqnarray}
For negative $\xi$ we could replace $\exp(-\xi_1/2)$ by $(1+|\xi_1|)^2$ where appears, but for our purpose this is not needed.

All the products in $\rho^{(m)}(\xi_1,\ldots,\xi_m)$ contain at least one $\exp(-\xi_i/2)$ for each $i$. In fact, this holds if: $K_{T;2,2}^{\textrm{edge}}(\xi_1,\xi_2)$ is not multiplied by
$K_{T;1,2}^{\textrm{edge}}(\xi_3,\xi_2)$, $K_{T;1,2}^{\textrm{edge}}(\xi_3,\xi_1)$,
$K_{T;2,1}^{\textrm{edge}}(\xi_1,\xi_3)$, $K_{T;2,1}^{\textrm{edge}}(\xi_2,\xi_3)$, and if $K_{T;1,2}^{\textrm{edge}}(\xi_1,\xi_2)$ is not multiplied by $K_{T;2,1}^{\textrm{edge}}(\xi_2,\xi_3)$. This is already proven in Lemma~\ref{lemmatwice}.

Consequently,
\begin{eqnarray}
& &\E_T\Big(|\eta_T^{\rm sym}(\Id_{[-M,\infty)})|^m\Big)
=\int_{[-M,\infty)^m}\dx\xi_1\ldots\dx\xi_m \rho^{(m)}_T(\xi_1,\ldots,\xi_m) \nonumber \\
&\leq & (2m)^{m/2} \Big(\int_{[-M,\infty)}C \exp(-\xi/2)\dx\xi\Big)^m  = 2^m C^m e^{M m/2}(2m)^{m/2}
\end{eqnarray}
uniformly in $T$. The term $(2m)^{m/2}$ comes from the fact that the absolute value of a determinant of a $n\times n$ matrix with entries of absolute value not exceeding $1$ is bounded by $n^{n/2}$ (Hadamard bound). Finally resetting the constant as $C 2\sqrt{2}$ the lemma is proved.
\end{proof}

To prove Theorem~\ref{thmMain} we use Theorem~\ref{thm0}, Proposition~\ref{prop1}, and Lemma~\ref{lemma6.3}.

\begin{proofOF}{of Theorem~\ref{thmMain}} Let us denote by $x_j$, $j\leq 0$, the position of the $j^{\rm th}$ element of $\zeta_T^{\textrm{flat}}$ and $x_j^{\textrm{sym}}$, $j\leq 0$, the position of the $j^{\rm th}$ element of $\zeta_T^{\textrm{sym}}$. Then define $\xi_{j,T}$ and $\xi_{j,T}^{\textrm{sym}}$ by
\begin{equation}
x_j=2T+\xi_{j,T}T^{1/3}2^{-2/3},\quad x_j^{\textrm{sym}}=4T+\xi_{j,T}^{\textrm{sym}} (2T)^{1/3}.
\end{equation}
By Proposition~\ref{prop1}, $x_j-j=\frac12(x_j^{\textrm{sym}}-j)$, which implies
\begin{equation}
\xi_{j,T}=\xi_{j,T}^{\textrm{sym}}+\frac{j}{(2T)^{1/3}}.
\end{equation}
Let $f_1,\ldots,f_m$ be test functions of compact support and denote by $M_f>0$ the minimal value such that $f_j(x)=0$ if $|x|\geq M_f$, $j=1,\ldots,m$. Then
\begin{eqnarray}\label{eq6.18}
\E_T\Big(\prod_{k=1}^m\eta_T^{\rm flat}(f_k)\Big) &=&\E_T\Big(\prod_{k=1}^m\sum_{i\leq 0} f_k(\xi_{i,T}^{\textrm{sym}}+i/(2T)^{1/3})\Big) \\
&=&\E_T\Big(\sum_{i_1,\ldots,i_m\leq 0}\prod_{k=1}^m f_k(\xi_{i_k,T}^{\textrm{sym}}+i_k/(2T)^{1/3})\Big). \nonumber
\end{eqnarray}
We bound the $f_k$'s by their supremum times $\Id_{[-M_f,M_f]}$ as in (\ref{eq7.8}), then
\begin{equation}
|\textrm{r.h.s.\ of }(\ref{eq6.18})|\leq \E_T\Big( \prod_{j=1}^m\sum_{i\leq 0} \Id_{[-M_f,M_f]}(\xi_{i,T}^{\textrm{sym}}+i/ (2T)^{1/3})\Big)\prod_{j=1}^m \|f_j\|_{\infty},
\end{equation}
and, since $\Id_{[-M_f,M_f]}(\xi_{i,T}^{\textrm{sym}}+i/(2T)^{1/3})\leq\Id_{[-M_f,\infty)}(\xi_{i,T}^{\textrm{sym}}+i/ (2T)^{1/3})\leq \Id_{[-M_f,\infty)}(\xi_{i,T}^{\textrm{sym}})$, it follows that
\begin{equation}
|\textrm{r.h.s.\ of }(\ref{eq6.18})| \leq \E_T\Big(\prod_{j=1}^m \eta_T^{\rm sym}(\Id_{[-M_f,\infty)})\Big) \prod_{j=1}^m \|f_j\|_{\infty}
\end{equation}
which is uniformly bounded in $T$ from Lemma~\ref{lemma6.3}. Therefore by Fubini's theorem,
\begin{equation}\label{eq6.20}
\E_T\Big(\prod_{k=1}^m\eta_T^{\rm flat}(f_k)\Big) =
\sum_{i_1,\ldots,i_m\leq 0}\E_T\Big(\prod_{k=1}^m f_k(\xi_{i_k,T}^{\textrm{sym}}+i_k/(2T)^{1/3})\Big).
\end{equation}
Moreover, $f_k(\xi_{i_k,T}^{\textrm{sym}}+i_k/(2T)^{1/3})=
f_k(\xi_{i_k,T}^{\textrm{sym}})+f'_k(\tilde \xi_{i_k,T})i_k/(2T)^{1/3}$ for some $\tilde \xi_{i_k,T}\in [\xi_{i_k,T}^{\textrm{sym}}+i_k/(2T)^{1/3},\xi_{i_k,T}^{\textrm{sym}}]$. Therefore (\ref{eq6.20}) equals
\begin{equation}\label{eq6.21}
\sum_{i_1,\ldots,i_m\leq 0}\E_T\Big(\prod_{k=1}^m f_k(\xi_{i_k,T}^{\textrm{sym}})\Big) = \E_T\Big(\prod_{k=1}^m\eta_T^{\rm sym}(f_k)\Big)
\end{equation}
plus $2^m-1$ terms which contains some $f'_k(\tilde \xi_{i_k,T})i_k/(2T)^{1/3}$. Finally we have to show that these terms vanish as $T\to\infty$. First we bound the $f_k$'s and the $f'_k$'s by $\|f_k\|_\infty$ and $\|f'_k\|_\infty$ times $\Id_{[-M_f,M_f]}$. Therefore each of the $2^m-1$ terms is bounded by a
\begin{equation}\label{eq6.22}
\frac{1}{T^{|J|/3}}\prod_{k\in I}\|f_k\|_\infty\prod_{k\in J}\|f'_k\|_\infty
\sum_{i_1,\ldots,i_m\leq 0}\prod_{k\in J}^m|i_k| \E_T\Big(\prod_{k=1}^m \Id_{[-M_f,\infty)}(\xi_{i_k,T}^{\textrm{sym}})\Big)
\end{equation}
where $I$ and $J$ are subset of $\{1,\ldots,m\}$ with $I\cup J = \{1,\ldots,m\}$ and $J$ is non-empty.
Let $j_0=\min\{i_1,\ldots,i_m\}$, then
\begin{eqnarray}\label{eq7.25}
& &\E_T\Big(\prod_{k=1}^m \Id_{[-M_f,\infty)}(\xi_{i_k,T}^{\textrm{sym}})\Big) = \E_T\Big( \Id_{[-M_f,\infty)}(\xi_{j_0,T}^{\textrm{sym}})\Big) \nonumber \\
&=&\Pb_T\Big(\xi_{j_0,T}^{\textrm{sym}} \geq -M_f\Big)\leq \Pb_T\Big(\eta_T^{\rm sym}(\Id_{[-M_f,\infty)})\geq j_0\Big) \\
&\leq & \frac{\E_T\Big(|\eta_T^{\rm sym}(\Id_{[-M_f,\infty)})|^{3m}\Big)}{|j_0|^{3m}}
\leq \frac{\Or(C^{3m} e^{M_f 3m/2}(3m)^{3m/2})}{\prod_{k=1}^m |i_k|^3}, \nonumber
\end{eqnarray}
since $|j_0| \geq |i_k|$ for all $k+1,\ldots,m$. From (\ref{eq7.25}) it follows that (\ref{eq6.22}) is uniformly bounded in $T$ and vanishes as $T\to\infty$. We have then proved that, for all $f_1,\ldots,f_m$ smooth test functions of compact support, 
\begin{equation}
\lim_{T\to\infty}\E_T\Big(\prod_{k=1}^m\eta_T^{\rm flat}(f_k)\Big) =\lim_{T\to\infty}\E_T\Big(\prod_{k=1}^m\eta_T^{\rm sym}(f_k)\Big)=\E\Big(\prod_{k=1}^m\eta^{\rm GOE}(f_k)\Big),
\end{equation}
the last equality being Theorem~\ref{thm0}.
\end{proofOF}

\subsection*{Acknowledgments}
The author would like to thank Michael Pr\"ahofer and Herbert Spohn for discussions about the present work, Tomohiro Sasamoto for explanations on the growth model in half-space, Kurt Johansson for suggesting the problem, Jani Lukkarinen for discussions on technical questions, and J\'ozsef L\H orinczi for reading part of the manuscript. Thanks go also to the referees for the critical reading and the useful suggestions.

\appendix
\section{Appendices}
\subsection{Bounds on the inverse of $A$}\label{AppAAA}
Let us denote the finite matrix $A$ by $A_N$ and its inverse by $A_N^{-1}$. For the $N=\infty$ case we use the notations $A$ and $A^{-1}$. Let us denote $I=\{-2N+1,\ldots,0\}$ and $L=\{-N+1,\ldots,0\}$.
Using (\ref{eq5.6}) we have
\begin{equation}\label{eqAppA1}
|A_{i,j}| \leq 1+\frac12 \sum_{k\geq i}\sum_{l\geq j} \frac{\tilde T^{k-i}}{(k-i)!}\frac{\tilde T^{l-j}}{(l-j)!}
= 1 + \frac12 e^{2\tilde T}.
\end{equation}
To obtain some properties of $A^{-1}$, we first estimate $[e^{-\tilde T \alpha_{-1}} P_- e^{-\tilde T \alpha_1}]_{i,j}$.
\begin{eqnarray}\label{eqAppA2}
[e^{-\tilde T \alpha_{-1}} P_- e^{-\tilde T \alpha_1}]_{i,j} &=& \sum_{\max\{i,j\}\leq k \leq 0}\frac{(-\tilde T)^{k-i}}{(k-i)!}\frac{(-\tilde T)^{k-j}}{(k-j)!} \\
&=& \sum_{l\geq 0}\frac{\tilde T^{2l} (-\tilde T)^{|i-j|}}{l! (l+|i-j|)!}-\sum_{l > -\max\{i,j\}}\frac{\tilde T^{2l} (-\tilde T)^{|i-j|}}{l! (l+|i-j|)!}\nonumber \\
&=& (-1)^{|i-j|} I_{|i-j|}(2\tilde T)-\sum_{l > -\max\{i,j\}}\frac{\tilde T^{2l} (-\tilde T)^{|i-j|}}{l! (l+|i-j|)!}, \nonumber
\end{eqnarray}
where $I_k$ is the modified Bessel function $I$ of order $k$. From (\ref{eqAppA2}) and $(l+|i-j|)!\geq l! |i-j|!$ follows
\begin{equation}\label{eqAppA3}
\big| [e^{-\tilde T \alpha_{-1}} P_- e^{-\tilde T \alpha_1}]_{i,j} \big| \leq I_{0}(2\tilde T) \frac{\tilde T^{|i-j|}}{|i-j|!} \leq \frac{\tilde T^{|i-j|}}{|i-j|!} e^{2\tilde T},
\end{equation}
which implies
\begin{equation}\label{eqAppA4}
\big| A^{-1}_{i,j} \big| \leq 2 \frac{\tilde T^{|i-j|}}{|i-j|!} e^{2\tilde T} \leq c_1(\tilde T) e^{-\mu_2(\tilde T) |i-j|},
\end{equation}
for some constants $c_1,\mu_2>0$.

The remainder sum in (\ref{eqAppA2}) is exponentially small in $-\max\{i,j\}$. In fact, for $n=-\max\{i,j\}$, 
\begin{eqnarray}\label{eqAppA5}
& &\big| [e^{-\tilde T \alpha_{-1}} P_- e^{-\tilde T \alpha_1}]_{i,j}-(-1)^{|i-j|} I_{|i-j|}(2\tilde T)\big|\nonumber \\& \leq& \frac{\tilde T^{|i-j|}}{|i-j|!}\sum_{l> n} \frac{\tilde T^{2l}}{(l!)^2}
\leq \frac{\tilde T^{|i-j|}}{|i-j|!} I_0(2\tilde T) e^{-\mu_1(\tilde T) n}
\end{eqnarray}
for some constant $\mu_1>0$. Thus, for all $(i,j)$ such that $\max\{i,j\} \leq -N/2$,
\begin{equation}\label{eqAppA6}
\big| A^{-1}_{i,j}-\lim_{m\to\infty} A^{-1}_{i-m,j-m}\big| \leq c_2(\tilde T) e^{-\mu_1 N/2}
\end{equation}
for some constant $c_2>0$, that is, in this region $A^{-1}$ is exponentially close to a Toepliz matrix.

For $j\in L$, using (\ref{eqAppA1}) and (\ref{eqAppA4}), we obtain
\begin{equation}\label{eqAppA7}
\big|[A_N A^{-1} - \Id]_{i,j}\big|= \sum_{l\leq -2N} A_{i,l}A^{-1}_{l,j} \leq c_3(\tilde T) e^{-\mu_2 N}
\end{equation}
with $c_3>0$ a constant.

\subsection{Some bounds}
\begin{lemma}\label{LemmaBessel1}
For $N\geq 0$,
\begin{equation}\label{eq5.2b}
|T^{1/3}J_{[2T+N T^{1/3}]}(2T)| \leq \exp(-N/2)\Or(1)
\end{equation}
uniformly in $T\geq T_0$ for some constant $T_0$.\\
For $N\leq 0$ it follows from a result of Landau~\cite{Lan00}, see (\ref{Landau}), that
\begin{equation}\label{eq5.2c}
|T^{1/3}J_{[2T+N T^{1/3}]}(2T)| \leq C
\end{equation}
uniformly in $T$ for a constant $C>0$.
\end{lemma}
\begin{proof} 
To obtain the bound we use 9.3.35 of~\cite{AS84}, i.e., for $z\in [0,1]$,
\begin{equation}\label{eq5.23}
J_{n}(n z)=\left(\frac{4\zeta}{1-z^2}\right)^{1/4} \left[ \frac{\Ai(n^{2/3} \zeta)}{n^{1/3}}(1+\Or(n^{-2}))+\frac{\Aip(n^{2/3} \zeta)}{n^{5/3}}\Or(1)\right]
\end{equation}
where
\begin{equation}\label{eq5.24}
\zeta(z)=(3/2)^{2/3}\left[\ln(1+\sqrt{1-z^2})-\ln(z)-\sqrt{1-z^2}\right]^{2/3}.
\end{equation}
In our case, $n=2T+N T^{1/3}$ and $z=(1+\e)^{-1}$ with $\e=\tfrac12 N T^{-2/3}\geq 0$. This implies that $z\in [0,1]$. In this interval the function $\zeta(z)$ is positive and decreasing. The prefactor is estimated using $4 \zeta(z(\e))(1-z(\e)^2)^{-1}2^{-4/3} \leq 1+\frac45 \e$ for all $\e>0$. Moreover, for $x\geq 0$, $\Ai(x)\leq\Ai(x/2)$ and $|\Aip(x)|\leq\Ai(x/2)$. Therefore
\begin{equation}\label{eq5.25}
|T^{1/3}J_{[2T+N T^{1/3}]}(2T)|\leq\left(1+\tfrac45 \e\right)^{1/4} \Ai(n^{2/3} \zeta/2)(1+\Or(T^{-4/3}))
\end{equation}
where we also used $(2T)^{1/3} \leq n^{1/3}$.
Next we bound (\ref{eq5.25}) separately for $N\leq \tfrac12 T^{2/3}$ and $N\geq \tfrac12 T^{2/3}$. 

\vspace{6pt}
\noindent Case 1) $0 \leq N \leq \tfrac12 T^{2/3}$. In this case $\e \leq \tfrac14$ and, for $\e\in [0,1/3]$, $\zeta(z(\e)) \geq \e$ holds. Replacing $n$ by $2T$ in the Airy function we have an upper bound since it is a decreasing function, consequently
\begin{equation}\label{eq5.26}
|T^{1/3}J_{[2T+N T^{1/3}]}(2T)| \leq 2 \Ai(N 2^{-4/3})(1+\Or(T^{-4/3})).
\end{equation}
Finally it is easy to verify that $2 \Ai(N 2^{-4/3}) \leq \exp(-N/2)$, and obtain the bound of the lemma.

\vspace{6pt}
\noindent Case 2) $N \geq \frac12 T^{2/3}$. In this case $\e\geq \frac14$ and $z(\e)\leq \tfrac45$. In this interval $\zeta(z) \geq \frac14 (\ln(8\e))^{2/3}$ from which follows
\begin{equation}\label{eq5.27}
|T^{1/3}J_{[2T+N T^{1/3}]}(2T)| \leq (N T^{-2/3})^{1/4}\Ai\left(\tfrac18 (n\ln(4N T^{-2/3}))^{2/3}\right)\Or(1).
\end{equation}
For $x\geq 0$, $\Ai(x)\leq \exp(-\tfrac23 x^{3/2})$, and $N \geq \frac12 T^{2/3}$ implies $\tilde N=4 N T^{-2/3} \geq 2$. Consequently,
\begin{eqnarray}\label{eq5.28}
|T^{1/3}J_{[2T+N T^{1/3}]}(2T)| &\leq& \tilde N^{1/4}\exp(-c_1 T (1+\tilde N/8))\Or(1)\nonumber \\
&\leq&\exp(-c_1 T)\exp(-2 c_2 T \tilde N) \tilde N^{1/4}\Or(1)
\end{eqnarray}
with $c_1 = \ln(2)/3$, $c_2=c_1/16$. For $T\geq 10$ and $\tilde N\geq 2$, $\tilde N^{1/4}\exp(-c_2 T \tilde N)\leq 1$, and $\exp(-c_2 T \tilde N) \leq \exp(-N/2)$ for $T$ large enough. These two last inequalities imply
\begin{equation}\label{eq5.29}
|T^{1/3}J_{[2T+N T^{1/3}]}(2T)| \leq \exp(-c_1 T)\exp(-N/2)\Or(1)
\end{equation}
for $T$ large enough, and the lemma is proved.
\end{proof}

\begin{lemma}\label{LemmaBessel2}
For all $N\geq 0$,
\begin{equation}\label{eq5.4b}
D_{T,N}=|T^{2/3}(J_{[2T+N T^{1/3}+1]}(2T)-J_{[2T+N T^{1/3}]}(2T))| \leq \exp(-N/2)\Or(1)
\end{equation}
uniformly in $T\geq T_0$ for some constant $T_0$.\\
For $N\leq 0$, there is a constant $C>0$ such that
\begin{equation}\label{eq5.4c}
D_{T,N}\leq C (1+|N|)
\end{equation}
uniformly in $T\geq 1$.
\end{lemma}
\begin{proof} First we consider $N\geq 0$.
 Let $N'=N+T^{-1/3}$, then we have to subtract $J_{[2T+N T^{1/3}]}(2T)$ to $J_{[2T+N' T^{1/3}]}(2T)$. In term of $\e=\tfrac12 N T^{-2/3}$ the difference is $1/(2T)$. Let us define
\begin{equation}
q(\e)=\left(\frac{4\zeta(z(\e))}{1-z(\e)^2}\right)^{1/4}(1+\e)^{-1/3},\quad p(\e)=(1+\e)^{2/3}\zeta(z(\e)),
\end{equation}
and
\begin{equation}
f(\e)=\frac{q(\e)}{(2T)^{1/3}}\Ai[(2T)^{2/3}p(\e)].
\end{equation}
With these notations,
\begin{eqnarray}
J_{[2T+N T^{1/3}]}(2T)&=&f(\e)+\frac{q(\e)}{(2T)^{1/3}}\Ai[(2T)^{2/3}p(\e)]\Or(T^{-2})\nonumber \\
& &+\frac{q(\e)}{(2T)^{1/3}}\Aip[(2T)^{2/3}p(\e)]\Or(T^{-4/3}).
\end{eqnarray}
Now we bound $D_{T,N}$ as follows.

\vspace{6pt}
\noindent Case 1) Let us consider $N\in[0,\frac12 T^{2/3}]$.
The second and the third terms are simply bounded by their absolute value. Then
\begin{eqnarray}\label{eq5.14b}
|D_{T,N}|&\leq& T^{2/3}\Big|f(\e+\frac{1}{2T})-f(\e)\Big|\nonumber \\
& &+ T^{2/3} \max_{x\in\{\e,\e+1/2T\}}\frac{q(x)}{(2T)^{1/3}}\Ai[(2T)^{2/3}p(x)]\Or(T^{-2}) \\
& &+T^{2/3}\max_{x\in\{\e,\e+1/2T\}} \frac{q(x)}{(2T)^{1/3}}|\Aip[(2T)^{2/3}p(x)]|\Or(T^{-4/3}).\nonumber
\end{eqnarray}
The first term is bounded by
\begin{equation}
T^{2/3}\Big|f(\e+\frac{1}{2T})-f(\e)\Big| \leq T^{2/3}\sup_{x\in[\e,\e+1/2T]}\big|f'(x)\big| \frac{1}{2T},
\end{equation}
where
\begin{equation}
\big|f'(x)\big| \leq |q'(x)|\frac{|\Ai[(2T)^{2/3}p(x)]|}{(2T)^{1/3}}+|q(x)p'(x)|\Ai[(2T)^{2/3}p(x)](2T)^{1/3}.
\end{equation}
We are considering the case of $N\in[0,\frac12 T^{2/3}]$, which corresponds to $\e\in [0,1/4]$. 
The functions $q$, $q'$, and $q\cdot p'$ behave modestly in this interval. They satisfy
\begin{equation}
q(x)\in [1.22,1.26],\quad |q'(x)| \in [0.14,0.17],\quad |q(x)p'(x)|\in [1.3,1.6]
\end{equation}
for $x\in[0,1/4]$. The Airy function and its derivative are bounded as in Lemma~\ref{LemmaBessel1}. Therefore
\begin{equation}
|D_{T,N}|\leq \exp(-N/2)(1+\Or(T^{-2/3})).
\end{equation}

\vspace{6pt}
\noindent Case 2) Let us consider $N\geq \frac12 T^{2/3}$. This case is simpler. We apply (\ref{eq5.29}) and obtain the bound
\begin{equation}
|D_{T,N}|\leq  T^{2/3} \exp(-c_1 T)\exp(-N/2)\Or(1) \leq \exp(-N/2)\Or(1)
\end{equation}
for $T$ large enough.

Secondly we consider $N\leq 0$. For $|N| \geq T^{1/3}$, using (\ref{Landau}) we obtain
\begin{equation}
|D_{T,N}|\leq c_3 T^{1/3} \leq c_3 |N|
\end{equation}
for some constant $c_3>0$. Next we consider $|N|\leq T^{1/3}$. Since $N$ is negative, $z\geq 1$ and (\ref{eq5.23}) holds with $\zeta(z)$ given by~\cite{AS84}
\begin{equation}
\zeta(z) =-(3/2)^{2/3}\left[\sqrt{z^2-1}-\arccos(1/z)\right]^{2/3}.
\end{equation}
Recall that $z=(1+\e)^{-1}$ and $\e = \frac12 N T^{-2/3}$. $|\e|\leq \frac12 T^{-1/3}$ is very close to zero. The estimate follows the same outline as for the case 1) for positive $N$. Take $T\geq 1$, then $\e\in[-\frac12,0]$ and 
\begin{equation}
q(\e)\in [1.25,1.37],\quad |q'(\e)| \in [0.16,0.25],\quad |q(\e)p'(\e)|\in [1.5,3.1].
\end{equation}
The difference is that now the Airy function in not rapidly decreasing since $p(\e)\leq 0$ and its derivative is even increasing. We use some simple bounds: $|\Ai(x)|\leq 1$ and $|\Aip(x)|\leq 1+|x|$ for all $x$, with the result
\begin{equation}
|D_{T,N}|\leq c_4 (1+|N|)(1+\Or(T^{-2/3}))
\end{equation}
for a constant $c_4>0$.
\end{proof}

\subsection{Some relations involving Bessel functions}
Here we give some relation on Bessel function~\cite{AS84} which are used in the work. Bessel functions $J_n$ are defined via the generating function by
\begin{equation}
\exp\left(\tfrac12 z (t-1/t)\right) = \sum_{k\in\Z} t^k J_k(z),\quad (t\neq 0).
\end{equation}
Then
\begin{enumerate}
\item for $n\in \N$, $J_{-n}(z)=(-1)^n J_n(z)$,
\item $J_0(z)+2\sum_{k\geq 1}J_{2k}(z)=1$,
\item $J_0^2(z)+2\sum_{k\geq 1}J_{k}^2(z)=1$,
\item for $n\geq 1$, $\sum_{k=0}^{2n} (-1)^k J_k(z) J_{2n-k}(z)+2\sum_{k=1}^\infty J_k(z) J_{2n+k}(z)=0$.
\end{enumerate}
Moreover the limit
\begin{equation}\label{Bessel2Airy}
\lim_{T\to\infty}T^{1/3}J_{[2T+\xi T^{1/3}]}(2T)=\Ai(\xi)
\end{equation}
holds. A useful result of Landau~\cite{Lan00} is the following:
\begin{equation}\label{Landau}
|J_n(x)| \leq c |x|^{-1/3}, \quad c=0.785...\textrm{ for all $n\in \Z$}.
\end{equation}

%\bibliographystyle{../../Biblio/patplain}
%\bibliography{../../Biblio/Biblio}
\end{document}